\documentclass[structabstract]{aa}  
\usepackage{natbib}
\usepackage{graphicx}
\usepackage[varg]{txfonts}
\usepackage{longtable}

\begin{document}
\title{Lithium abundances along the RGB: FLAMES-GIRAFFE spectra of a large
  sample of low-mass Bulge stars\thanks{Based on observations at the Very Large
    Telescope of the European Southern Observatory, Cerro Paranal/Chile under
    Programme 083.D-0046(A). Table~\ref{onltab} is only available in electronic
    form at the CDS via anonymous ftp to cdsarc.u-strasbg.fr (130.79.128.5) or
    via http://cdsweb.u-strasbg.fr/cgi-bin/qcat?J/A+A/.}}
\author{
  T. Lebzelter\inst{1}
  \and
  S. Uttenthaler\inst{1}\fnmsep\thanks{The first two authors have contributed
    equally to this paper.}
  \and
  M. Busso\inst{2}
  \and
  M. Schultheis\inst{3}
  \and
  B. Aringer\inst{4}
}

\institute{
  University of Vienna, Department of Astronomy, T\"urkenschanzstra\ss e 17,
  1180 Vienna, Austria\\
  \email{thomas.lebzelter;stefan.uttenthaler@univie.ac.at}
  \and
  Dipartimento di Fisica, Universit\`{a} di Perugia, and INFN, Sezione di
  Perugia. Via Pascoli, 06123 Perugia, Italy
  \and
  Observatoire de Besan\c{c}on, 41bis, avenue de l'Observatoire,
  25000 Besan\c{c}on, France
  \and
  INAF - Padova Astronomical Observatory, Vicolo dell'Osservatorio 5, 35122,
  Padova, Italy
}

\date{Received July 21, 2011; accepted November 07, 2011}

 
\abstract
    {
      A small number of K-type giants on the red giant branch (RGB) is known to
      be very rich in lithium (Li). This fact is not accounted for by standard
      stellar evolution theory. The exact phase and mechanism of Li enrichment
      is still a matter of debate.
}
    {Our goal is to probe the abundance of Li along the RGB, from its base to
      the tip, to confine Li-rich phases that are supposed to occur on the RGB.}
    {For this end, we obtained medium-resolution spectra with the FLAMES
      spectrograph at the VLT in GIRAFFE mode for a large sample of 401
      low-mass RGB stars located in the Galactic bulge. The Li abundance was
      measured in the stars with a detectable Li 670.8\,nm line by means of
      spectral synthesis with COMARCS model atmospheres. A new 2MASS
      $(J-K_S) - T_{\rm{eff}}$ calibration from COMARCS models is presented
      in the Appendix.}
    {Thirty-one stars with a detectable Li line were identified, three of which
      are Li-rich according to the usual criterion
      ($\log\epsilon({\rm Li})>1.5$). The stars are distributed all along the
      RGB, not concentrated in any particular phase of the red giant evolution
      (e.g.\ the luminosity bump or the red clump). The three Li-rich stars are
      clearly brighter than the luminosity bump and red clump, and do not show
      any signs of enhanced mass loss.}
    {We conclude that the Li enrichment mechanism cannot be restricted to a
      clearly defined phase of the RGB evolution of low-mass stars
      ($M\sim1M_{\sun}$), contrary to earlier suggestions from disk field stars.
}
    
    \keywords{Stars: late-type -- Stars: evolution -- Stars: abundances}
\authorrunning{Lebzelter et al.}
\titlerunning{Lithium in Bulge RGB stars}
   \maketitle

\section{Introduction}

Lithium (Li) is an important diagnostic tool in stellar evolution because its
abundance strongly depends on the ambient conditions. It is quickly destroyed
at $T > 3 \times 10^6$\,K so that it diminishes if the stellar surface is
brought into contact with hot layers by mixing processes. However, if the
overturn time scale for mixing becomes faster than the decay of the parent
$^7$Be, then Li destruction reverts into production through the so-called
Cameron-Fowler mechanism \citep{CF71}. In low-mass stars during the
main-sequence phase, the initial Li abundance strongly decreases, hence slow
mixing should prevail \citep{Mic86}. During the ascent on the red giant branch
(RGB), any Li remaining in the envelope is further diluted by the first
dredge-up (FDU); stellar models excluding atomic diffusion and rotation predict
a surface Li abundance at FDU of $\log\epsilon({\rm Li}) \le +1.5$ (where
$\log\epsilon({\rm Li}) = \log[N({\rm Li})/N({\rm H})]+12$) at this stage. This
is indeed what is observed in most G-K giants \citep{Lam80,Bro89,Mis06}, where
some of these stars show Li even far below the expectations \citep{Mal99}.
However, about 1 -- 2\,\% of the K giants have a high Li abundance and are
therefore called {\em Li-rich} \citep{Bro89,dlR97}.

Advanced evolutionary stages in which Li abundances might possibly increase
through quite fast mixing have been identified by
\citet[][hereafter CB00]{CB00}, in the moments at which the discontinuity of
molecular weight $\mu$ left behind by the downward envelope expansion is erased
by the advancement of the H-burning shell. This occurs at the bump of the
luminosity function, which is on the RGB for low masses. For intermediate mass
stars, this moment is postponed until after core He-burning, when the star
climbs for the second time towards the Hayashi track and the envelope deepens
again (on the so-called asymptotic giant branch, or AGB). The absence of a
$\mu$-barrier is the key factor because it allows any diffusive or plume-like
mixing mechanism to link the envelope with inner layers, and in some cases does
it fast enough, to activate a Cameron-Fowler mechanism for Li production.
However, the proposed location of the enrichment phase directly at the RGB bump
has been questioned by recent observational results \citep[e.g.][]{Mon11}.
Observations of Li in metal-poor globular cluster stars show that a slow
extra-mixing episode at the RGB bump efficiently destroys any Li remaining
after FDU \citep{Lind09b}. Hence, evolved Li-rich stars brighter than the RGB
bump are evidence that those stars somehow skip this mixing episode or even
undergo a phase of Li production, which motivates the search for these stars. As
it is unlikely that they skipped Li destruction at FDU, they must have somehow
replenished it through Be decay \citep[see the discussion in][]{Palm11}.

While the Li abundance has been measured already in an increasing number of
field stars \citep[e.g.][]{Mal99,Kumar11,Mon11} the samples investigated always
had the disadvantage of either inhomogeneity, especially in mass and age,
uncertainty of the distance, or strong selection biases (e.g.\ only stars
around the RGB bump). This circumstance hampers drawing definite conclusions on
the evolutionary phase and length of the Li enrichment. Homogeneous samples,
such as RGB stars in a cluster, are small \citep[e.g.][]{Pas01,Pas04}. We
therefore designed a programme to spectroscopically observe a large and
homogeneous sample of stars from the bottom to the tip of the RGB belonging
to the Galactic bulge (GB) to derive their Li abundances. The GB offers a large
number of low-mass giants at roughly equal distance within a small area in the
sky, perfectly suited for observations with a multi-object spectrograph such as
FLAMES-GIRAFFE. Furthermore, in contrast to globular clusters, the Bulge
contains stars in a wide range of metallicities, which provides the opportunity
to check for any dependence of the Li enrichment processes on this parameter.
We present the results of this programme concerning our search for Li. It will
not only help to confine the phases and duration of Li production on the RGB,
but will also provide a reference value for the Li abundance in the outer GB,
which is needed for interpreting the first Li-rich AGB stars detected there
recently \citep[][hereafter U07]{Utt07}. The present data set is also of high
interest for the study of the structure and nature of the bulge itself, the
results of which will be presented in a forthcoming paper (Uttenthaler et al.,
in preparation).


\section{Sample and observations}\label{obs}

\subsection{Target selection}

\begin{figure}
  \centering
  \includegraphics[width=\linewidth,bb=87 366 538 700,clip]{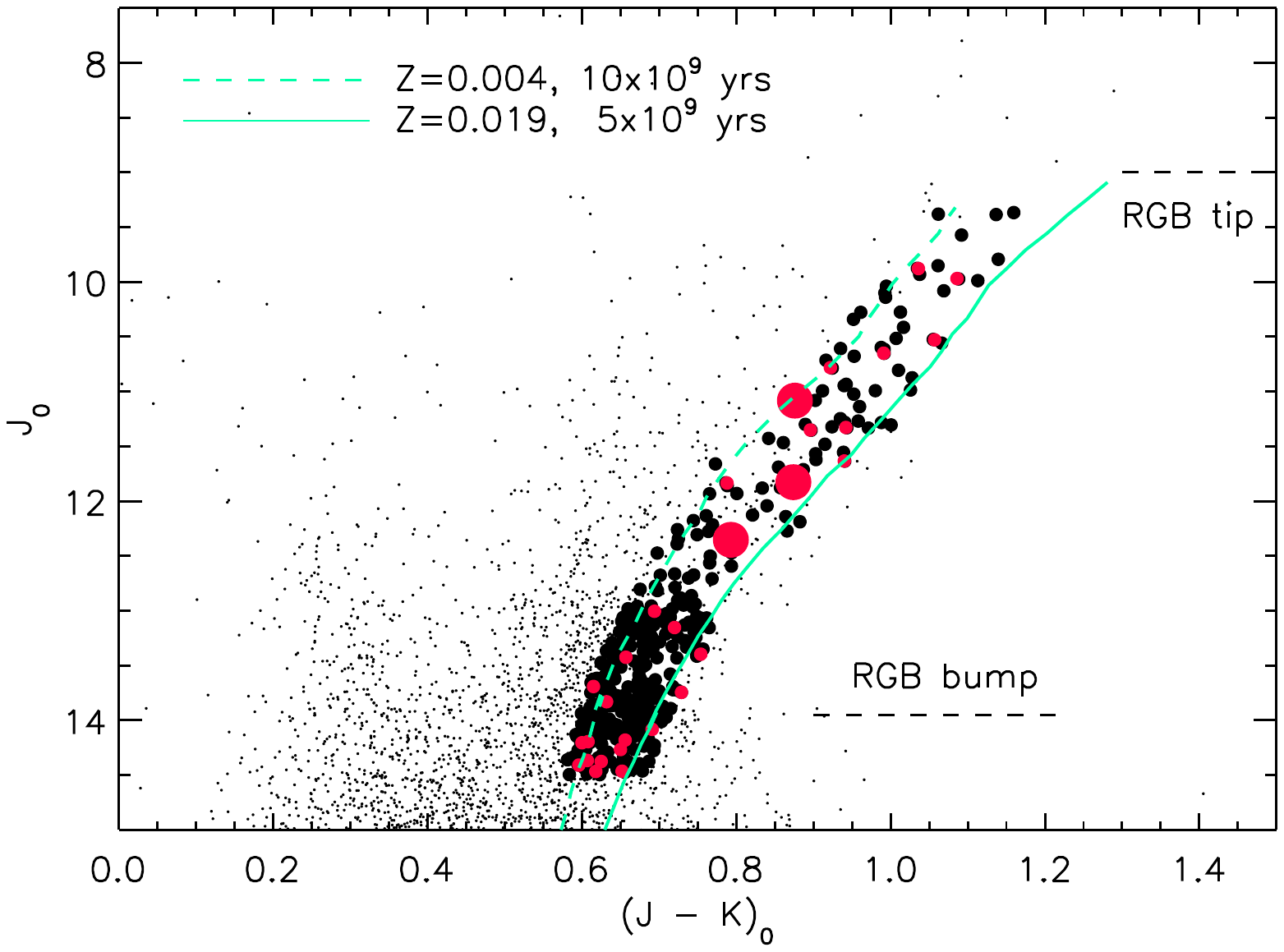}
  \caption{2MASS Colour-magnitude diagram of the field around
    $(l,b)=(0\degr,-10\degr)$. Stars not observed in the present programme are
    plotted as black dots, observed targets as black circles, stars moderately
    enriched in Li as small red circles, and Li-rich stars as big red circles.
    Two RGB isochrones from \citet{Gir00}, with metallicities and ages as
    indicated in the legend, are also included. The isochrones are truncated at
    the tip of the RGB.}
  \label{cmd}
\end{figure}

The selection of targets was based on data from the 2MASS catalogue
\citep{2MASS} in a 25\arcmin\ diameter circle towards the direction
$(l,b)=(0\degr,-10\degr)$, which is the centre of the Palomar-Groningen field
no.~3 (PG3). This field in the outer GB was chosen because the sample of AGB
stars studied by U07 is also located in the PG3. The GB stars in this field are
roughly 1.4\,kpc away from the Galactic plane and centre. A colour-magnitude
diagram (CMD) to illustrate the target selection is displayed in
Fig.~\ref{cmd}. The observed targets (black circles) were chosen to fall close
to two isochrones from \citet{Gir00}, which are representative for the GB:
$Z=0.004$ and age $10\times10^9$ years, and $Z=0.019$
\citep[which is $Z_{\odot}$ on the scale used by][]{Gir00} and age $5\times10^9$
years. A distance modulus of $14\fm5$ to the GB was adopted. The chosen targets
were allowed to have a $(J - K)_0$ colour either bracketed by the isochrones,
or $0\fm02$ redder or bluer than these. We dereddened the stars using the
linear relation for the reddening in the $B_J$ photographic band as a function
of galactic latitude given in \citet{Schulthe98}. To translate this into the
extinction in the $J$- and $K$-bands, we used the relation $R = A_{V}/E(B- V)$
with $R = 3.2$ and the reddening law of \citet{GS2003}. We find a mean
reddening of $0\fm129$ in the $J$-band and $0\fm049$ in the $K$-band. A
slightly higher reddening is found using the map of \citet{Schleg98}, but the
difference is within the errors of the 2MASS photometry. In addition to the
selection with the colour criterion we selected only stars fainter than
$J_0=9\fm0$ to exclude AGB stars above the tip of the RGB, and stars brighter
than $J_0=14\fm5$ to include the RGB bump, which is expected from isochrones at
$13\fm8 \leqslant J_0 \leqslant 14\fm1$. Furthermore, we discarded all targets
that had fewer than two quality flags 'A' in the 2MASS $JHK$ photometry and
targets that had another source within 3\arcsec\ that was not fainter by at
least $2\fm0$ in the $J$-band than the target itself. This selection scheme
provides a bias against close visual pairs in our sample. Applying all
those criteria yielded 514 targets for the observations.

Also discernible in Fig.~\ref{cmd} are the two red clumps (RCs) recently
identified by \citet{Nat10} and \citet{MZ10}, which probably represent two
populations at different distance in the GB, which are of similar age and
metallicity \citep{DeP11}. These structures were not known to us in the design
phase of the programme. Initially, the fainter RC at $J_0 \sim 13\fm9$ was
interpreted by us to be the RGB bump, and only the brighter one at
$J_0 \sim 13\fm2$ as the red clump. This recent discovery introduces an
uncertainty of a bulge star's location on the RGB of about 0\fm7. The locations
of the RGB bumps of the two populations are then expected to be slightly above
and below the theoretical line marked in Fig.~\ref{cmd}. Given this range of
distance moduli ($\sim14\fm1$ to $\sim14\fm8$) and the photometric errors from
the 2MASS catalogue, we can estimate the expected range of metallicities in our
sample with the help of the isochrones. We find that at the lower brightness
end, stars down to a metallicity of $[{\rm M}/{\rm H}]=-1.3$ would still fall
in the selection range, albeit with a lower probability than stars with a
metallicity between that of the two selection isochrones. At brighter
magnitudes, the lower metallicity cut will be somewhat higher, but stars down to
$[{\rm M}/{\rm H}]=-0.7$ should be complete in the whole magnitude range of our
sample. On the other hand, stars belonging to the near side of the X-shaped
bulge may still fall within our selection region up to a metallicity of
$[{\rm M}/{\rm H}]=+0.2$. As we will see in Sect.~\ref{analysis}, a few stars
fall outside this range, but those cases should be rare.

\subsection{Observations and data reduction}

To measure the Li abundance we obtained spectra of the sample stars using the
FLAMES spectrograph in the GIRAFFE configuration. The grating HR15 centred at
665\,nm was used, which gave a spectral resolution of 17\,000 for a wavelength
coverage from 644 to 682\,nm. This setting contains the Li resonance line at
670.8\,nm and the H$\alpha$ line, in addition to several other atomic and
molecular features. Because our sample stretches over a brightness range of 5
magnitudes, it was divided into three sub-groups, a bright
(\mbox{$11\fm9 \lesssim R \lesssim 13\fm3$}), an intermediate
(\mbox{$13\fm3 \lesssim R \lesssim 14\fm7$}), and a faint group
(\mbox{$14\fm7 \lesssim R \lesssim 16\fm2$}), with total exposure times
of 0.5, 1, and 4\,h, respectively\footnote{The $R$-band limits were determined
from the isochrone predictions and used in the GIRAFFE exposure time
calculator.}. For the bright group one fibre configuration was sufficient, while
the larger number of targets at fainter magnitudes made necessary to split up
the intermediate and faint group into two and three fibre configuration,
respectively.
The observations were obtained in service mode in June 2009. To keep observing
blocks below one hour and to handle cosmic ray hits, the observations of the
intermediate group were split into two, those of the faint group into five
exposures. Individual spectra were extracted from each of these images
separately using the FLAMES pipeline provided by ESO. The quality of each
spectrum was checked, and those with low signal or cosmic ray hits in the
critical wavelength ranges were rejected. Before averaging the good spectra,
small wavelength shifts caused by splitting the individual observations in some
cases over a few weeks were corrected. The applied shift agreed with the
expected value for the difference in heliocentric correction, and the same
shift was found for all spectra obtained at a given time. Therefore, we can
exclude that these shifts are the result of orbital motion in a binary system.
No other velocity shifts were found within the accuracy limits of our data.
However, for several stars only one observation was available, therefore we
point out that with the present data set an effective discrimination against
binaries is not possible. We aimed at a signal-to-noise ratio (S/N) close to
100, which was achieved for most targets. Of the initially proposed sample, 401
spectra had sufficiently high quality to check for the presence of the Li line
and to measure their radial velocity (RV).

\subsection{Foreground contamination}

An important point for this kind of study is the foreground contamination of
the sample. To reliably estimate the fraction of genuine bulge stars in our
sample, we ran two simulations of Galactic population models, namely the
Besan\c{c}on model \citep{Rob03} and the TRILEGAL model \citep{Gir05}. For the
TRILEGAL simulation (kindly provided by E.\ Vanhollebeke), the best-fit
parameters of a tri-axial bulge found by \citet{Van09} were adopted. We applied
the same colour and magnitude selection criteria to the simulated data as to
the observed data. This selection yielded 81\% and 74\% bulge stars in the
TRILEGAL and Besan\c{c}on model, respectively. We also inspected the
distribution of initial masses of the simulated stars in the selection area.
The mass spectrum in the TRILEGAL simulation is relatively sharply peaked at
$1.01 M_{\sun}$, with a standard deviation of $0.13 M_{\sun}$. Hence, we are
confident that our sample indeed contains mostly low-mass ($M \sim 1 M_{\sun}$)
red giant stars that are located in the GB.

Some of the foreground stars might be identified by their high proper motion.
We therefore searched the Southern Proper Motion Program catalogue version 4
\citep[SPM4;][]{Gir11} for sample stars with high proper motion. All but ten of
our sample stars were identified in the SPM4 catalogue. Unfortunately, the
uncertainties in the SPM4 catalogue are large, so that bulge and foreground
stars can hardly be separated. We decided to assign those stars to the
foreground whose proper motion is higher than 20\,mas/yr. Applying this
criterion, we found 50 foreground star candidates; these stars are marked with
an asterisk in Table~\ref{onltab}. One star in our sample (\#300) has a
particularly high proper motion of more than 200\,mas/yr, hence it is probably
a nearby K dwarf.

\section{Analysis and results}\label{analysis}

For the following analysis we used hydrostatic COMARCS atmospheres in
connection with the COMA spectral synthesis package \citep{Ari09}. This ensures
a consistent treatment of radiative transfer and opacities during the complete
analysis. We started with a temperature calibration based on 2MASS $(J-K_S)_0$
colours, which is presented in Appendix~\ref{JKTeffsect}. A determination of
the stellar temperature and surface gravity using only the FLAMES spectra was
not possible because of the limited spectral range and the dominance of TiO
lines in the cooler targets. After fixing the temperature based on the colours,
we obtained the surface gravities $(g)$ from isochrones in Fig.~\ref{cmd}
\citep{Gir00}, assuming that the stars belong to the Bulge RGB. The uncertainty
in the distance modulus of $\pm0\fm35$ (Sect.~\ref{obs}) introduces an
uncertainty in $\log(g)$ of $\pm0.16$, which has only a very minor impact on
the uncertainty in the Li abundance, however.


In a first step of the analysis of the spectra, we determined the RVs of the
stars by cross-correlating the observed spectra with a synthetic spectrum that
was based on a COMARCS model with a temperature close to that estimated for
each star. In the next step, the observed spectra were inspected visually
around the 670.8\,nm Li line, together with a synthetic spectrum calculated
with no Li to check for excess absorption. The Li line was found to be
detectable in 31 stars. Four of these Li-bearing stars are also foreground star
candidates (Sect.~\ref{obs}).

To estimate the metallicity of each star, we calculated model atmospheres with
five different metallicities ($[{\rm M}/{\rm H}]=-1.5$, $-1.0$, $-0.5$, $0.0$,
and $+0.5$), with $T_{{\rm eff}}$ and $\log g$ values determined in the previous
step. A general $\alpha$-element enhancement of $[\alpha/{\rm Fe}]=+0.2$, a C/O
ratio of 0.3, and a micro-turbulent velocity typical for red giant stars of
$\xi=2.5$\,km\,s$^{-1}$ were assumed. The spectra with the five different
metallicities were interpolated and fitted to the observed spectra in the
wavelength range 649 -- 680\,nm using the downhill simplex IDL routine
amoeba.pro. The aim of this exercise was not to determine a very precise
metallicity of the stars, but instead to reasonably model the quasi-continuum in
the vicinity of the Li line, which is suppressed from the true continuum by a
forest of weak lines, and the lines blending with the Li line. This procedure
resulted in very satisfying fits of the spectra. The metallicity determined in
this way is probably quite reliable, its accuracy is mainly limited by the
accuracy of the temperature determination. A check of this method on the
observed spectrum of Arcturus yields a metallicity of
$[{\rm M}/{\rm H}]=-0.69$. However, if model atmospheres without an
alpha-enhancement are used, this would be increased to $-0.64$, which
reasonably agrees with the range of metallicities found for this star in the
Simbad database. We find a mean metallicity of $[{\rm M}/{\rm H}]=-0.48$ for
the 31 Li-bearing stars, which is somewhat lower than what would be expected
from previous investigations of Bulge fields close to ours
\citep{Zoc08}\footnote{Performing the averaging on a {\em linear} scale, which
is the mathematically correct way, the mean is $[{\rm M}/{\rm H}]=-0.21$.}.
Some of the stars have very weak metal lines, indeed.
Possibly, there are selection effects introduced in the sense that the Li-rich
phenomenon could be more common among metal-poor stars, and/or that the
detection threshold of Li is lower at lower metallicity because of the reduced
background of blending lines. Note that the three most Li-rich stars (see below)
have a metallicity between $-0.83$ and $-0.96$\,dex below solar.

With the stellar parameters fixed in this way, we synthesised spectra, assuming
varying Li abundances to fit the observed spectra. Atomic line data for the
spectral synthesis were taken from the VALD data base \citep{Kup99}. The line
list was checked by comparing a model spectrum of Arcturus
\citep[adopting the stellar parameters and abundances found by][]{Ryd10} with
the observed Arcturus spectrum \citep{Hin00} in the vicinity of the 671\,nm Li
line. Some lines present in VALD were found to be too strong in the model
spectrum; their $\log(gf)$ were decreased accordingly, or the line was
completely removed from the list. The abundances given in \citet{Caf09} were
adopted as solar reference. We did not apply NLTE corrections to the Li
abundances \citep{Lind09a} derived under LTE assumption because these
corrections are usually small and, for the parameters covered by our sample
stars, within the uncertainties.

\begin{figure}
  \centering
  \includegraphics[width=\linewidth,bb=82 369 538 699,clip]{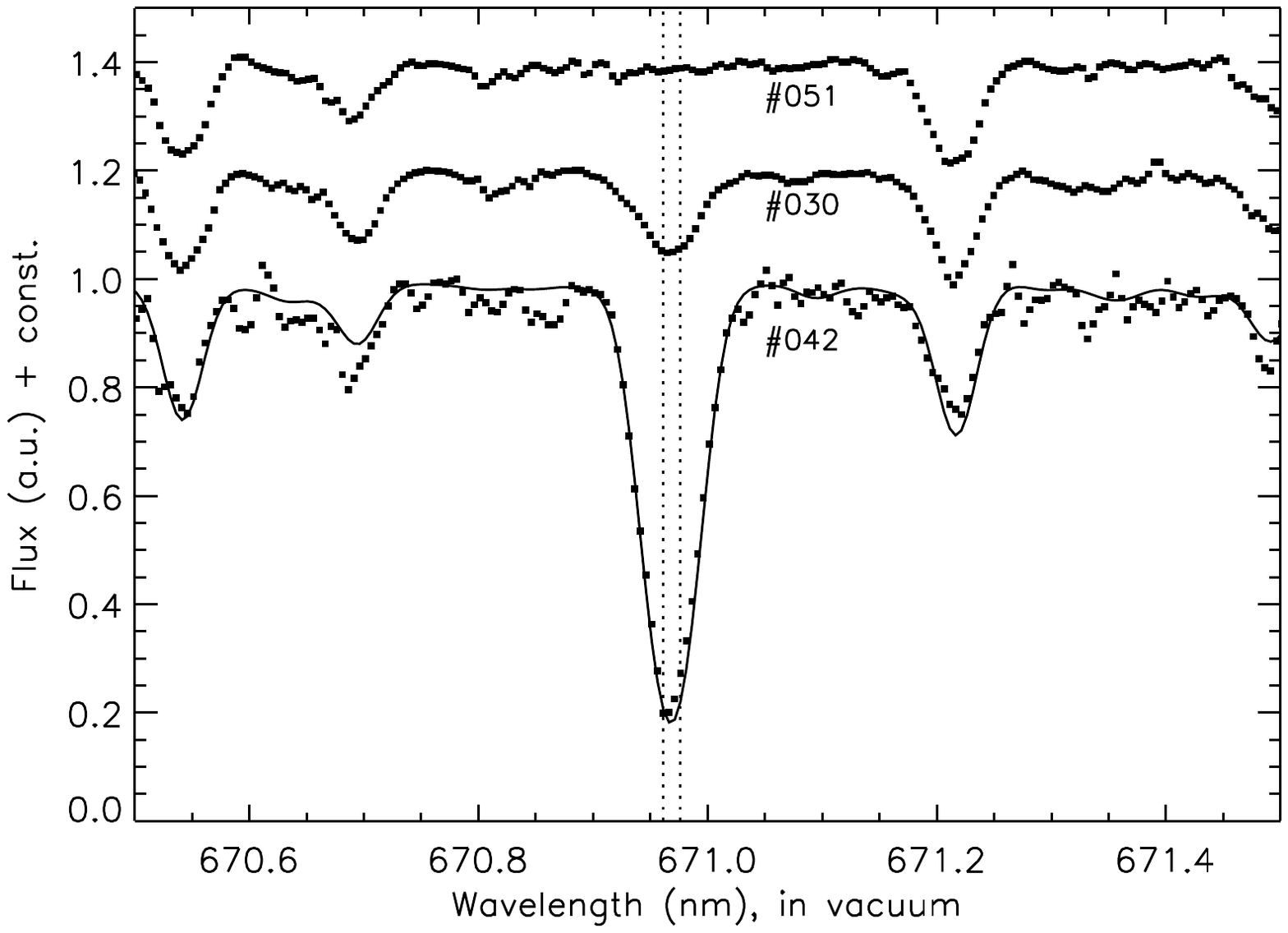}
  \caption{Observed spectra of the the programme stars \#042, the most Li-rich
    star in our sample; \#030, a Li-detected star; and \#051, a Li-poor star
    (black dots, from bottom to top). For clarity, the spectra of stars \#030
    and \#051 have been shifted by +0.2 and +0.4 in flux. The continuous line
    is the best-fit synthetic spectrum to star \#042 with a Li abundance of
    $\log \epsilon({\rm Li})=+3.2$. The two dotted vertical lines indicate the
    laboratory wavelengths of the hyperfine transitions in $^7$Li. Note also
    the good fit to two Fe lines at \mbox{$\sim670.55$} and $\sim671.21$\,nm,
    which indicates a well-determined me\-tal\-li\-ci\-ty of
    $[{\rm M}/{\rm H}]=-0.85$ for star \#042. The metallicities for the two
    other stars were determined to be $-0.94$ (\#030) and $-1.03$ (\#051),
    respectively.
  }
  \label{tatltuae}
\end{figure}

The programme stars \#042, \#080, and \#123 are identified to be Li-rich
according to the classical definition, one of them (\#042) has a Li abundance
that agrees with the cosmic value and should be regarded as {\em super}
Li-rich. The observed spectrum of star \#042 around the Li line together with
a synthetic spectrum with the best fitting abundances is displayed in
Fig.~\ref{tatltuae}. The spectrum of a Li-detected star (\#030,
$\log\epsilon({\rm Li})=+0.1$) and of a Li-poor star (\#051) are also shown in
that figure. All other stars with a detected Li line have an abundance at or
below the solar abundance. The error in the Li abundance was estimated by
varying the stellar parameters by their uncertainty. The temperature has the
highest influence on the strength of the Li line, and is uncertain by
$\pm160$\,K for the hotter stars, and $\pm90$\,K for the cooler stars, based on
the uncertainty in the 2MASS photometry. We consequently estimate an
uncertainty in the Li abundance of $\pm0.3$\,dex. Moreover, the detection
threshold varies with the temperature because Li becomes more and more ionised
at higher temperature. A detection threshold of $\log\epsilon({\rm Li})=+0.5$
was derived for the hottest sample stars (4800\,K), which declines to 0.0 at
around 4200\,K, and to $\sim-0.5$ at 3600\,K\footnote{At still lower
temperatures, the detection threshold rises again because of the increasing
strength of the TiO lines.}.

All measured Li abundances as well as important quantities of the targets are
collected in Table~\ref{onltab}. Our results are also illustrated in
Fig.~\ref{cmd}: Stars with a detected Li line are plotted as small red circles,
and the Li-rich stars are plotted as large red circles.

\onllongtab{1}{
\begin{longtable}{lcccccrccccc}
\caption{Parameters of the stars and measured abundances.}\label{onltab}\\
\hline
\hline
No.\ & RA(J2000) & Dec(J2000) & $J$ & $H$ & $K$ & RV          & $T_{\rm eff}$ & $\log g$ & [M/H] & $\log\epsilon({\rm Li})$ & FG\\
            &  deg.\  & deg.\    &     &     &     & km\,s$^{-1}$ & K          &          &        &   \\
 (1) & (2) & (3) & (4) & (5) & (6) & (7) & (8) & (9) & (10) & (11) & (12) \\
\hline
\endfirsthead
\caption{Continued.}\\
\hline
\hline
No.\ & RA(J2000) & Dec(J2000) & $J$ & $H$ & $K$ & RV          & $T_{\rm eff}$ & $\log g$ & [M/H] & $\log\epsilon({\rm Li})$ & FG\\
            &  deg.\  & deg.\    &     &     &     & km\,s$^{-1}$ & K          &          &        &   \\
 (1) & (2) & (3) & (4) & (5) & (6) & (7) & (8) & (9) & (10) & (11) & (12) \\
\hline
\endhead
\hline
\endfoot
\hline
\endlastfoot
001 & 276.814513 & $-33.589966$ &  9.495 &  8.524 &  8.256 &   117.328  & 3465 &       &       &       & *\\
002 & 276.544710 & $-33.787083$ &  9.512 &  8.624 &  8.369 &    20.277  & 3697 &       &       &       &  \\
003 & 276.608107 & $-33.667065$ &  9.516 &  8.615 &  8.298 &    76.484  & 3520 &       &       &       &  \\
005 & 276.918684 & $-33.751503$ &  9.696 &  8.820 &  8.527 &    55.090  & 3629 &       &       &       &  \\
006 & 276.432104 & $-33.784958$ &  9.926 &  9.061 &  8.704 & $ -29.587$ & 3514 &       &       &       &  \\
007 & 276.714893 & $-33.891556$ &  9.978 &  9.096 &  8.838 &    12.913  & 3698 &       &       &       &  \\
008 & 276.600416 & $-33.681347$ & 10.007 &  9.121 &  8.891 & $-135.108$ & 3756 &  0.5  &$-0.14$&  0.4  &  \\
009 & 276.807769 & $-33.757442$ & 10.057 &  9.178 &  8.941 &   106.494  & 3749 &       &       &       &  \\
010 & 276.873765 & $-33.726830$ & 10.098 &  9.183 &  8.932 & $-128.768$ & 3637 &  0.6  &$-0.59$&$-0.3$ &  \\
011 & 276.817925 & $-33.649303$ & 10.115 &  9.191 &  8.923 & $ -59.147$ & 3578 &       &       &       &  \\
012 & 276.611791 & $-33.631435$ & 10.170 &  9.330 &  9.094 &    74.703  & 3843 &       &       &       &  \\
013 & 276.506457 & $-33.723846$ & 10.213 &  9.322 &  9.062 &    27.664  & 3681 &       &       &       &  \\
015 & 276.488172 & $-33.844830$ & 10.232 &  9.400 &  9.158 &    85.083  & 3847 &       &       &       &  \\
016 & 276.682337 & $-33.955391$ & 10.266 &  9.421 &  9.194 &    25.704  & 3845 &       &       &       &  \\
017 & 276.595604 & $-33.694199$ & 10.406 &  9.574 &  9.312 &    50.134  & 3803 &       &       &       &  \\
018 & 276.592533 & $-33.625336$ & 10.409 &  9.579 &  9.366 & $ -59.013$ & 3914 &       &       &       &  \\
019 & 276.586978 & $-33.867764$ & 10.470 &  9.639 &  9.438 &   178.014  & 3934 &       &       &       &  \\
020 & 276.704314 & $-33.794868$ & 10.542 &  9.678 &  9.446 &    33.422  & 3794 &       &       &       & *\\
021 & 276.653399 & $-33.955818$ & 10.642 &  9.747 &  9.556 &    72.019  & 3815 &       &       &       &  \\
022 & 276.471245 & $-33.828197$ & 10.657 &  9.769 &  9.520 & $ -80.548$ & 3711 &  0.9  &$+0.08$&  0.5  & *\\
023 & 276.565931 & $-33.660713$ & 10.690 &  9.821 &  9.542 & $ -42.688$ & 3687 &       &       &       &  \\
024 & 276.667873 & $-33.869152$ & 10.726 &  9.949 &  9.659 & $  -7.078$ & 3857 &       &       &       &  \\
025 & 276.771326 & $-33.748791$ & 10.735 &  9.924 &  9.721 & $-131.548$ & 3969 &       &       &       &  \\
026 & 276.566599 & $-33.576355$ & 10.747 &  9.916 &  9.673 &    15.044  & 3849 &       &       &       &  \\
027 & 276.621730 & $-33.925117$ & 10.779 &  9.943 &  9.708 & $  -0.464$ & 3849 &  0.9  &$+0.11$&  0.6  &  \\
028 & 276.566492 & $-33.894100$ & 10.807 & 10.005 &  9.774 & $ -78.865$ & 3932 &       &       &       &  \\
029 & 276.632910 & $-33.850189$ & 10.843 & 10.032 &  9.847 &   305.574  & 4008 &       &       &       &  \\
030 & 276.876718 & $-33.749680$ & 10.911 & 10.121 &  9.909 &    13.784  & 3991 &  1.0  &$-0.94$&  0.1  &  \\
031 & 276.833702 & $-33.646965$ & 10.933 & 10.095 &  9.844 & $ -12.983$ & 3808 &       &       &       &  \\
032 & 276.664170 & $-33.889168$ & 11.002 & 10.132 &  9.895 &    59.999  & 3770 &       &       &       &  \\
033 & 276.608510 & $-33.775723$ & 11.064 & 10.240 & 10.041 &    46.289  & 3953 &       &       &       &  \\
034 & 276.530697 & $-33.808578$ & 11.077 & 10.243 & 10.056 &    72.605  & 3959 &       &       &       &  \\
035 & 276.734398 & $-33.905167$ & 11.112 & 10.288 & 10.008 &    37.840  & 3775 &       &       &       &  \\
036 & 276.693013 & $-33.893147$ & 11.119 & 10.247 & 10.060 &   110.060  & 3874 &       &       &       &  \\
037 & 276.455229 & $-33.789265$ & 11.126 & 10.329 & 10.132 &    25.919  & 4017 &       &       &       &  \\
038 & 276.531016 & $-33.587563$ & 11.157 & 10.323 & 10.122 & $-105.342$ & 3933 &       &       &       &  \\
039 & 276.476724 & $-33.652061$ & 11.191 & 10.461 & 10.223 & $ -99.668$ & 4074 &  1.2  &$-0.17$&  0.3  &  \\
040 & 276.632713 & $-33.608341$ & 11.204 & 10.397 & 10.238 & $-213.279$ & 4075 &       &       &       &  \\
041 & 276.611649 & $-33.623035$ & 11.212 & 10.433 & 10.228 &     7.052  & 4037 &       &       &       &  \\
042 & 276.514543 & $-33.867790$ & 11.212 & 10.396 & 10.256 &     3.906  & 4096 &  1.2  &$-0.85$&  3.2  &  \\
043 & 276.545896 & $-33.910049$ & 11.236 & 10.429 & 10.287 & $ -60.971$ & 4109 &  1.2  &$-0.61$&  0.3  &  \\
044 & 276.501519 & $-33.712631$ & 11.270 & 10.456 & 10.228 & $ -66.401$ & 3917 &       &       &       &  \\
045 & 276.497877 & $-33.881786$ & 11.378 & 10.535 & 10.362 &   240.911  & 3969 &       &       &       &  \\
046 & 276.469519 & $-33.651062$ & 11.403 & 10.619 & 10.362 & $  -1.719$ & 3921 &       &       &       &  \\
047 & 276.718276 & $-33.890141$ & 11.407 & 10.593 & 10.388 &    29.551  & 3957 &       &       &       &  \\
048 & 276.593535 & $-33.598366$ & 11.415 & 10.584 & 10.345 & $  -9.884$ & 3857 &       &       &       &  \\
049 & 276.751404 & $-33.686954$ & 11.428 & 10.661 & 10.459 &    81.687  & 4065 &       &       &       &  \\
050 & 276.668148 & $-33.844818$ & 11.433 & 10.578 & 10.353 &    29.158  & 3829 &       &       &       &  \\
051 & 276.897946 & $-33.802887$ & 11.446 & 10.612 & 10.445 & $ -53.729$ & 3992 &       &       &       &  \\
052 & 276.827522 & $-33.672203$ & 11.457 & 10.650 & 10.435 & $ -28.859$ & 3952 &  1.4  &$+0.09$&  0.4  &  \\
053 & 276.598353 & $-33.763283$ & 11.464 & 10.658 & 10.412 &    16.198  & 3893 &       &       &       &  \\
054 & 276.753347 & $-33.748871$ & 11.480 & 10.655 & 10.504 &    21.622  & 4049 &  1.2  &$-0.94$&       &  \\
058 & 276.763338 & $-33.726757$ & 11.555 & 10.775 & 10.634 & $-133.305$ & 4167 &       &       &       &  \\
060 & 276.814847 & $-33.799526$ & 11.592 & 10.814 & 10.653 &    19.204  & 4126 &       &       &       &  \\
061 & 276.745468 & $-33.614552$ & 11.610 & 10.791 & 10.615 & $-206.421$ & 4011 &       &       &       &  \\
065 & 276.767964 & $-33.723507$ & 11.682 & 10.895 & 10.664 & $ -59.417$ & 3961 &       &       &       &  \\
067 & 276.764757 & $-33.714439$ & 11.695 & 10.914 & 10.713 & $ -42.029$ & 4036 &       &       &       &  \\
069 & 276.732436 & $-33.691238$ & 11.750 & 10.953 & 10.767 &    28.948  & 4035 &       &       &       &  \\
070 & 276.809016 & $-33.594540$ & 11.763 & 10.955 & 10.743 & $ -54.899$ & 3957 &  1.4  &$-0.39$&  0.2  &  \\
072 & 276.594486 & $-33.730743$ & 11.791 & 11.111 & 10.937 & $ -52.206$ & 4325 &       &       &       &  \\
073 & 276.836739 & $-33.717419$ & 11.816 & 11.071 & 10.883 & $ -56.535$ & 4139 &       &       &       &  \\
075 & 276.910260 & $-33.767544$ & 11.837 & 11.034 & 10.873 &    45.375  & 4070 &       &       &       & *\\
080 & 276.663676 & $-33.680828$ & 11.956 & 11.185 & 11.002 &    67.611  & 4099 &  1.5  &$-0.83$&  2.0  &  \\
081 & 276.598738 & $-33.673889$ & 11.964 & 11.282 & 11.096 & $-155.581$ & 4294 &  1.5  &$-0.32$&  0.7  &  \\
084 & 276.723840 & $-33.562702$ & 11.989 & 11.339 & 11.120 &    29.832  & 4290 &       &       &       &  \\
085 & 276.663499 & $-33.799038$ & 12.006 & 11.199 & 11.069 &    58.329  & 4134 &       &       &       & *\\
086 & 276.810910 & $-33.617214$ & 12.009 & 11.269 & 11.096 &     9.035  & 4185 &       &       &       &  \\
088 & 276.843442 & $-33.804676$ & 12.058 & 11.340 & 11.215 &    84.524  & 4343 &       &       &       &  \\
089 & 276.708411 & $-33.645660$ & 12.060 & 11.331 & 11.179 & $ -58.044$ & 4261 &       &       &       &  \\
096 & 276.728929 & $-33.625156$ & 12.172 & 11.457 & 11.252 &    38.482  & 4171 &       &       &       &  \\
101 & 276.650041 & $-33.653530$ & 12.257 & 11.553 & 11.355 & $-123.588$ & 4213 &       &       &       &  \\
102 & 276.603598 & $-33.725780$ & 12.263 & 11.594 & 11.421 & $ -37.926$ & 4352 &       &       &       &  \\
104 & 276.764294 & $-33.611069$ & 12.270 & 11.528 & 11.326 &    23.454  & 4119 &       &       &       &  \\
107 & 276.796933 & $-33.752850$ & 12.304 & 11.596 & 11.481 &    16.073  & 4392 &       &       &       &  \\
108 & 276.818554 & $-33.721001$ & 12.315 & 11.546 & 11.354 & $ -55.377$ & 4080 &       &       &       &  \\
110 & 276.831116 & $-33.661152$ & 12.343 & 11.625 & 11.495 & $-109.396$ & 4334 &       &       &       &  \\
112 & 276.844085 & $-33.719276$ & 12.386 & 11.681 & 11.584 & $  -3.301$ & 4444 &       &       &       &  \\
115 & 276.876631 & $-33.669006$ & 12.398 & 11.613 & 11.454 & $-172.477$ & 4115 &  1.7  &$-0.56$&$<-0.3$&  \\
116 & 276.886129 & $-33.817890$ & 12.401 & 11.716 & 11.560 & $ -54.145$ & 4346 &       &       &       & *\\
117 & 276.839217 & $-33.735954$ & 12.407 & 11.626 & 11.535 & $-141.307$ & 4277 &       &       &       &  \\
118 & 276.605428 & $-33.607613$ & 12.438 & 11.750 & 11.607 &    31.103  & 4380 &  1.7  &$-1.19$&$<-0.3$&  \\
121 & 276.850407 & $-33.727764$ & 12.470 & 11.850 & 11.667 & $  -2.194$ & 4441 &       &       &       &  \\
123 & 276.733228 & $-33.686802$ & 12.482 & 11.722 & 11.609 & $ -85.450$ & 4278 &  1.7  &$-0.96$&  2.1  &  \\
128 & 276.674814 & $-33.794342$ & 12.520 & 11.823 & 11.717 &    32.514  & 4446 &       &       &       &  \\
131 & 276.862902 & $-33.842964$ & 12.596 & 11.871 & 11.725 & $ -28.810$ & 4277 &       &       &       &  \\
132 & 276.529376 & $-33.635666$ & 12.607 & 11.953 & 11.827 & $-159.866$ & 4514 &       &       &       &  \\
133 & 276.754996 & $-33.640835$ & 12.630 & 11.933 & 11.784 &   136.734  & 4341 &       &       &       & *\\
136 & 276.670396 & $-33.678337$ & 12.694 & 11.991 & 11.848 & $ -16.510$ & 4342 &       &       &       &  \\
139 & 276.669257 & $-33.560745$ & 12.724 & 12.024 & 11.849 & $  -8.143$ & 4278 &       &       &       &  \\
142 & 276.592785 & $-33.609268$ & 12.796 & 12.171 & 11.994 & $-120.627$ & 4454 &       &       &       &  \\
143 & 276.720586 & $-33.791100$ & 12.801 & 12.219 & 11.977 & $ -28.694$ & 4391 &       &       &       & *\\
144 & 276.724865 & $-33.593685$ & 12.806 & 12.156 & 12.024 &    47.662  & 4504 &       &       &       &  \\
147 & 276.605327 & $-33.599644$ & 12.834 & 12.130 & 12.014 & $ -13.128$ & 4407 &       &       &       &  \\
148 & 276.808190 & $-33.749828$ & 12.835 & 12.127 & 11.988 & $ -40.138$ & 4335 &       &       &       &  \\
154 & 276.884497 & $-33.711712$ & 12.903 & 12.231 & 12.130 & $ -74.863$ & 4521 &       &       &       &  \\
156 & 276.802245 & $-33.615543$ & 12.915 & 12.209 & 12.115 &     5.379  & 4453 &       &       &       & *\\
158 & 276.607949 & $-33.626446$ & 12.936 & 12.308 & 12.179 &    47.731  & 4574 &       &       &       &  \\
159 & 276.506642 & $-33.859463$ & 12.947 & 12.320 & 12.168 & $ -67.344$ & 4514 &       &       &       &  \\
167 & 276.696311 & $-33.687943$ & 12.992 & 12.354 & 12.170 &    16.036  & 4398 &       &       &       &  \\
169 & 276.585971 & $-33.583626$ & 13.016 & 12.356 & 12.206 & $ -21.109$ & 4434 &       &       &       &  \\
172 & 276.674085 & $-33.722809$ & 13.036 & 12.399 & 12.232 &    62.310  & 4444 &       &       &       &  \\
178 & 276.673132 & $-33.740860$ & 13.069 & 12.451 & 12.242 & $ -14.891$ & 4385 &       &       &       &  \\
180 & 276.715472 & $-33.633411$ & 13.078 & 12.449 & 12.266 &    57.926  & 4424 &       &       &       & *\\
182 & 276.772886 & $-33.712326$ & 13.085 & 12.412 & 12.316 & $ -44.163$ & 4535 &       &       &       &  \\
183 & 276.791120 & $-33.629635$ & 13.088 & 12.478 & 12.337 & $ -74.273$ & 4584 &       &       &       &  \\
184 & 276.841313 & $-33.728474$ & 13.093 & 12.437 & 12.271 &    13.471  & 4393 &       &       &       &  \\
186 & 276.721044 & $-33.780018$ & 13.110 & 12.489 & 12.371 &    29.773  & 4618 &       &       &       &  \\
187 & 276.533345 & $-33.873089$ & 13.112 & 12.435 & 12.314 & $-142.788$ & 4462 &       &       &       &  \\
188 & 276.753476 & $-33.945801$ & 13.119 & 12.543 & 12.370 &    19.007  & 4585 &       &       &       &  \\
190 & 276.538801 & $-33.644936$ & 13.134 & 12.484 & 12.360 &    11.192  & 4530 &  2.1  &$+0.27$&  0.5  &  \\
191 & 276.627097 & $-33.841202$ & 13.146 & 12.522 & 12.389 & $-175.576$ & 4569 &       &       &       &  \\
192 & 276.790756 & $-33.846645$ & 13.151 & 12.493 & 12.363 & $  -8.106$ & 4481 &       &       &       &  \\
193 & 276.717695 & $-33.835640$ & 13.153 & 12.548 & 12.410 &    56.213  & 4605 &       &       &       &  \\
194 & 276.864580 & $-33.723576$ & 13.154 & 12.547 & 12.417 & $ -33.430$ & 4620 &       &       &       &  \\
195 & 276.694217 & $-33.601109$ & 13.163 & 12.508 & 12.393 & $  -3.184$ & 4537 &       &       &       &  \\
196 & 276.841761 & $-33.689163$ & 13.167 & 12.539 & 12.385 & $ -44.337$ & 4499 &       &       &       & *\\
197 & 276.749744 & $-33.856068$ & 13.171 & 12.572 & 12.406 & $ -22.102$ & 4544 &       &       &       &  \\
198 & 276.638867 & $-33.930489$ & 13.179 & 12.505 & 12.351 &    89.036  & 4381 &       &       &       &  \\
199 & 276.601595 & $-33.643906$ & 13.190 & 12.594 & 12.394 &    55.281  & 4469 &       &       &       &  \\
200 & 276.910018 & $-33.774948$ & 13.190 & 12.585 & 12.420 & $ -23.265$ & 4527 &       &       &       &  \\
201 & 276.655418 & $-33.667728$ & 13.192 & 12.617 & 12.458 & $ -12.106$ & 4637 &       &       &       &  \\
202 & 276.600953 & $-33.920620$ & 13.193 & 12.561 & 12.403 & $ -72.007$ & 4480 &       &       &       &  \\
203 & 276.718361 & $-33.917198$ & 13.194 & 12.505 & 12.362 & $-112.531$ & 4370 &       &       &       &  \\
204 & 276.651481 & $-33.734238$ & 13.196 & 12.496 & 12.354 & $ -44.013$ & 4351 &       &       &       &  \\
205 & 276.778831 & $-33.720421$ & 13.203 & 12.613 & 12.452 & $ -66.094$ & 4583 &       &       &       &  \\
206 & 276.529719 & $-33.594994$ & 13.210 & 12.613 & 12.466 &    26.825  & 4613 &       &       &       &  \\
207 & 276.854131 & $-33.731907$ & 13.221 & 12.595 & 12.447 &    62.408  & 4519 &       &       &       &  \\
208 & 276.628729 & $-33.743752$ & 13.226 & 12.621 & 12.494 & $ -39.333$ & 4642 &       &       &       &  \\
209 & 276.798530 & $-33.662918$ & 13.233 & 12.540 & 12.415 &    95.619  & 4405 &       &       &       &  \\
210 & 276.836967 & $-33.769405$ & 13.243 & 12.643 & 12.456 &    21.124  & 4484 &       &       &       &  \\
211 & 276.756279 & $-33.788357$ & 13.248 & 12.575 & 12.430 & $ -34.348$ & 4404 &       &       &       &  \\
212 & 276.892143 & $-33.753590$ & 13.248 & 12.670 & 12.487 &    21.790  & 4552 &       &       &       & *\\
215 & 276.537753 & $-33.603622$ & 13.256 & 12.649 & 12.491 & $-119.873$ & 4555 &       &       &       & *\\
216 & 276.630558 & $-33.724087$ & 13.258 & 12.652 & 12.484 &    38.871  & 4526 &       &       &       &  \\
217 & 276.611185 & $-33.719536$ & 13.258 & 12.650 & 12.506 &    53.834  & 4586 &       &       &       &  \\
218 & 276.533293 & $-33.793808$ & 13.263 & 12.630 & 12.449 &    25.411  & 4421 &       &       &       &  \\
219 & 276.490017 & $-33.791950$ & 13.267 & 12.645 & 12.524 &    70.196  & 4614 &       &       &       &  \\
220 & 276.786346 & $-33.598869$ & 13.267 & 12.685 & 12.496 &    36.370  & 4531 &       &       &       & *\\
221 & 276.488605 & $-33.676846$ & 13.277 & 12.662 & 12.508 & $ -68.765$ & 4544 &       &       &       &  \\
222 & 276.765113 & $-33.566185$ & 13.280 & 12.582 & 12.447 &    89.637  & 4371 &       &       &       &  \\
223 & 276.689571 & $-33.850536$ & 13.283 & 12.596 & 12.483 &    63.721  & 4452 &  2.1  &$-0.01$&  0.5  &  \\
224 & 276.530160 & $-33.772533$ & 13.283 & 12.630 & 12.546 & $-159.749$ & 4630 &       &       &       & *\\
225 & 276.490013 & $-33.854095$ & 13.287 & 12.654 & 12.528 & $  -2.227$ & 4568 &       &       &       &  \\
226 & 276.441487 & $-33.833378$ & 13.287 & 12.601 & 12.440 &    86.733  & 4344 &       &       &       &  \\
227 & 276.460830 & $-33.865948$ & 13.295 & 12.693 & 12.545 & $ -30.132$ & 4593 &       &       &       &  \\
228 & 276.567606 & $-33.881664$ & 13.305 & 12.729 & 12.573 &     5.293  & 4641 &       &       &       &  \\
229 & 276.488238 & $-33.744823$ & 13.308 & 12.659 & 12.507 & $ -93.491$ & 4457 &       &       &       &  \\
230 & 276.701957 & $-33.724052$ & 13.317 & 12.670 & 12.492 & $ -15.167$ & 4390 &       &       &       & *\\
231 & 276.521504 & $-33.882675$ & 13.318 & 12.634 & 12.517 &    87.725  & 4454 &       &       &       &  \\
232 & 276.701163 & $-33.812206$ & 13.321 & 12.791 & 12.593 &    46.717  & 4650 &       &       &       &  \\
233 & 276.742927 & $-33.655430$ & 13.327 & 12.768 & 12.577 & $ -33.817$ & 4588 &       &       &       &  \\
234 & 276.817331 & $-33.599476$ & 13.328 & 12.681 & 12.582 &     3.010  & 4598 &       &       &       &  \\
235 & 276.561797 & $-33.754608$ & 13.331 & 12.760 & 12.590 &     2.381  & 4618 &       &       &       &  \\
236 & 276.811041 & $-33.747356$ & 13.339 & 12.698 & 12.552 & $ -36.405$ & 4485 &       &       &       & *\\
237 & 276.543824 & $-33.638008$ & 13.349 & 12.771 & 12.601 & $-128.142$ & 4601 &       &       &       &  \\
238 & 276.707179 & $-33.863255$ & 13.363 & 12.721 & 12.560 & $ -48.485$ & 4443 &       &       &       &  \\
239 & 276.514484 & $-33.893902$ & 13.367 & 12.706 & 12.596 & $  -6.349$ & 4534 &       &       &       &  \\
240 & 276.494178 & $-33.865170$ & 13.374 & 12.671 & 12.545 & $ -45.410$ & 4384 &       &       &       &  \\
241 & 276.691228 & $-33.652210$ & 13.381 & 12.740 & 12.621 &   163.130  & 4563 &       &       &       &  \\
242 & 276.769980 & $-33.622349$ & 13.400 & 12.785 & 12.587 &    37.049  & 4420 &       &       &       &  \\
243 & 276.445539 & $-33.820004$ & 13.401 & 12.802 & 12.662 & $-201.762$ & 4626 &       &       &       & *\\
245 & 276.458400 & $-33.680145$ & 13.410 & 12.870 & 12.674 & $  -0.726$ & 4637 &       &       &       &  \\
246 & 276.732205 & $-33.735016$ & 13.413 & 12.722 & 12.633 & $ -20.965$ & 4506 &       &       &       &  \\
247 & 276.652760 & $-33.666142$ & 13.415 & 12.859 & 12.647 & $ -12.734$ & 4542 &       &       &       &  \\
248 & 276.663722 & $-33.563961$ & 13.423 & 12.782 & 12.703 & $ -91.605$ & 4680 &       &       &       &  \\
249 & 276.874535 & $-33.706005$ & 13.432 & 12.805 & 12.675 &    20.002  & 4564 &       &       &       &  \\
251 & 276.814218 & $-33.891129$ & 13.436 & 12.813 & 12.703 & $-100.233$ & 4630 &       &       &       &  \\
252 & 276.590433 & $-33.824047$ & 13.447 & 12.815 & 12.625 &    14.280  & 4399 &       &       &       &  \\
253 & 276.698884 & $-33.720577$ & 13.450 & 12.802 & 12.707 & $ -12.354$ & 4608 &       &       &       &  \\
254 & 276.560443 & $-33.786304$ & 13.457 & 12.905 & 12.659 &    47.473  & 4463 &       &       &       &  \\
256 & 276.885368 & $-33.847267$ & 13.459 & 12.843 & 12.729 &    47.665  & 4637 &       &       &       &  \\
257 & 276.673226 & $-33.739399$ & 13.464 & 12.896 & 12.744 &    53.227  & 4676 &       &       &       &  \\
258 & 276.509126 & $-33.638287$ & 13.466 & 12.795 & 12.648 &    20.249  & 4414 &       &       &       &  \\
259 & 276.903426 & $-33.794418$ & 13.475 & 12.812 & 12.641 &     0.338  & 4362 &       &       &       &  \\
260 & 276.856050 & $-33.636570$ & 13.477 & 12.791 & 12.647 & $ -99.944$ & 4375 &       &       &       &  \\
261 & 276.597652 & $-33.706051$ & 13.478 & 12.889 & 12.705 & $ -56.387$ & 4530 &       &       &       &  \\
262 & 276.627210 & $-33.776421$ & 13.481 & 12.854 & 12.717 & $ -31.979$ & 4552 &       &       &       &  \\
263 & 276.659752 & $-33.951805$ & 13.489 & 12.884 & 12.727 & $  -5.578$ & 4553 &       &       &       &  \\
264 & 276.545369 & $-33.932869$ & 13.493 & 12.880 & 12.781 & $ -61.467$ & 4701 &       &       &       &  \\
266 & 276.533760 & $-33.787659$ & 13.505 & 12.874 & 12.744 & $ -68.790$ & 4562 &       &       &       &  \\
267 & 276.616984 & $-33.586750$ & 13.506 & 12.954 & 12.792 & $-101.202$ & 4699 &       &       &       &  \\
268 & 276.494876 & $-33.647800$ & 13.515 & 12.919 & 12.794 &    71.564  & 4681 &       &       &       & *\\
269 & 276.764459 & $-33.589680$ & 13.518 & 12.940 & 12.784 &    69.289  & 4635 &       &       &       &  \\
270 & 276.614926 & $-33.938892$ & 13.526 & 12.948 & 12.808 &    65.396  & 4680 &       &       &       &  \\
271 & 276.465499 & $-33.869881$ & 13.526 & 12.857 & 12.692 &    29.413  & 4372 &  2.3  &$+0.48$&  0.6  &  \\
272 & 276.827432 & $-33.753338$ & 13.529 & 12.833 & 12.770 & $-104.732$ & 4559 &       &       &       &  \\
273 & 276.785331 & $-33.659843$ & 13.538 & 12.873 & 12.710 &   223.568  & 4381 &       &       &       &  \\
274 & 276.842104 & $-33.732456$ & 13.541 & 12.852 & 12.714 & $ -21.443$ & 4381 &       &       &       &  \\
275 & 276.765440 & $-33.581825$ & 13.554 & 12.999 & 12.817 &    68.660  & 4626 &  2.3  &$-0.76$&  1.0  &  \\
276 & 276.728640 & $-33.949757$ & 13.557 & 12.951 & 12.756 & $-121.711$ & 4446 &       &       &       &  \\
277 & 276.505794 & $-33.774723$ & 13.561 & 12.941 & 12.782 & $ -51.159$ & 4515 &       &       &       &  \\
278 & 276.714377 & $-33.724106$ & 13.581 & 13.007 & 12.860 &    16.094  & 4672 &       &       &       &  \\
279 & 276.619384 & $-33.610786$ & 13.591 & 12.968 & 12.842 &    50.290  & 4596 &       &       &       &  \\
280 & 276.699742 & $-33.801800$ & 13.600 & 13.064 & 12.896 & $ -22.286$ & 4722 &       &       &       &  \\
281 & 276.632739 & $-33.920406$ & 13.601 & 12.931 & 12.843 & $ -74.670$ & 4565 &       &       &       &  \\
282 & 276.603736 & $-33.792931$ & 13.606 & 13.010 & 12.886 & $ -59.064$ & 4678 &       &       &       &  \\
283 & 276.839527 & $-33.652874$ & 13.614 & 12.951 & 12.909 & $ -47.148$ & 4717 &       &       &       & *\\
284 & 276.724473 & $-33.882050$ & 13.619 & 12.995 & 12.858 & $  -6.884$ & 4555 &       &       &       &  \\
285 & 276.810256 & $-33.899017$ & 13.632 & 12.985 & 12.875 &    24.234  & 4562 &       &       &       & *\\
286 & 276.629095 & $-33.587997$ & 13.646 & 13.024 & 12.914 & $-104.407$ & 4645 &       &       &       &  \\
287 & 276.860156 & $-33.865948$ & 13.646 & 13.033 & 12.895 &    75.530  & 4578 &       &       &       &  \\
288 & 276.589341 & $-33.604183$ & 13.659 & 13.078 & 12.939 & $ -63.114$ & 4682 &       &       &       &  \\
291 & 276.787583 & $-33.871380$ & 13.673 & 13.065 & 12.954 & $ -22.181$ & 4673 &       &       &       &  \\
292 & 276.501441 & $-33.769341$ & 13.689 & 13.075 & 12.971 & $ -60.436$ & 4688 &       &       &       &  \\
294 & 276.807204 & $-33.744389$ & 13.705 & 13.002 & 12.912 & $ -92.853$ & 4469 &       &       &       &  \\
295 & 276.671937 & $-33.722317$ & 13.712 & 13.142 & 13.001 & $  -7.833$ & 4704 &       &       &       &  \\
296 & 276.472514 & $-33.777775$ & 13.719 & 13.094 & 12.963 &   121.918  & 4578 &       &       &       &  \\
297 & 276.686566 & $-33.655930$ & 13.744 & 13.148 & 13.021 & $-110.334$ & 4669 &       &       &       &  \\
298 & 276.665050 & $-33.818905$ & 13.746 & 13.132 & 13.029 &    87.153  & 4684 &       &       &       &  \\
299 & 276.859868 & $-33.871136$ & 13.748 & 13.085 & 12.995 & $-105.991$ & 4572 &       &       &       &  \\
300 & 276.442637 & $-33.752930$ & 13.750 & 13.103 & 13.048 & $ -23.401$ & 4737 &       &       &       & *\\
301 & 276.695536 & $-33.764545$ & 13.751 & 13.160 & 13.029 & $ -68.258$ & 4669 &       &       &       &  \\
302 & 276.749584 & $-33.841690$ & 13.752 & 13.130 & 13.059 & $-165.133$ & 4752 &       &       &       &  \\
304 & 276.677892 & $-33.890118$ & 13.758 & 13.194 & 12.984 &    54.041  & 4521 &       &       &       &  \\
305 & 276.676672 & $-33.567745$ & 13.771 & 13.127 & 13.063 & $ -54.614$ & 4715 &       &       &       &  \\
306 & 276.686629 & $-33.707558$ & 13.777 & 13.146 & 13.024 & $ -57.854$ & 4581 &       &       &       &  \\
307 & 276.712714 & $-33.944599$ & 13.782 & 13.237 & 13.092 & $ -73.813$ & 4760 &       &       &       & *\\
308 & 276.542295 & $-33.749485$ & 13.790 & 13.240 & 13.092 & $-103.250$ & 4746 &       &       &       &  \\
309 & 276.845057 & $-33.643093$ & 13.797 & 13.196 & 13.031 & $  -7.766$ & 4542 &       &       &       &  \\
310 & 276.737301 & $-33.623081$ & 13.801 & 13.204 & 13.097 &    68.140  & 4724 &       &       &       &  \\
311 & 276.855065 & $-33.892948$ & 13.813 & 13.230 & 13.106 & $ -10.062$ & 4706 &       &       &       &  \\
312 & 276.805007 & $-33.749069$ & 13.817 & 13.247 & 13.105 & $ -18.932$ & 4695 &       &       &       &  \\
313 & 276.803927 & $-33.834248$ & 13.821 & 13.261 & 13.126 &    29.782  & 4744 &  2.4  &$-0.71$&  0.6  & *\\
314 & 276.622601 & $-33.794971$ & 13.822 & 13.150 & 13.024 & $ -60.428$ & 4460 &       &       &       &  \\
315 & 276.821083 & $-33.773975$ & 13.822 & 13.218 & 13.084 &    96.113  & 4618 &       &       &       & *\\
316 & 276.689193 & $-33.710064$ & 13.845 & 13.207 & 13.090 &     0.896  & 4575 &       &       &       & *\\
317 & 276.527656 & $-33.798317$ & 13.846 & 13.287 & 13.138 & $ -20.465$ & 4716 &       &       &       &  \\
318 & 276.605035 & $-33.901566$ & 13.846 & 13.198 & 13.094 & $  -8.413$ & 4583 &       &       &       &  \\
319 & 276.676819 & $-33.581738$ & 13.852 & 13.225 & 13.053 & $ -58.955$ & 4460 &       &       &       &  \\
320 & 276.826131 & $-33.769146$ & 13.853 & 13.137 & 13.051 & $ -41.389$ & 4444 &       &       &       & *\\
321 & 276.547129 & $-33.646622$ & 13.857 & 13.271 & 13.155 &    11.187  & 4736 &       &       &       &  \\
322 & 276.625298 & $-33.608067$ & 13.858 & 13.294 & 13.140 & $ -54.255$ & 4687 &       &       &       &  \\
323 & 276.850420 & $-33.753113$ & 13.859 & 13.172 & 13.081 & $ -52.017$ & 4508 &       &       &       & *\\
324 & 276.648919 & $-33.838467$ & 13.868 & 13.295 & 13.114 &    29.848  & 4577 &       &       &       & *\\
325 & 276.887809 & $-33.831329$ & 13.871 & 13.297 & 13.117 & $ -84.110$ & 4569 &       &       &       & *\\
326 & 276.520313 & $-33.606983$ & 13.871 & 13.248 & 13.174 & $ -13.705$ & 4753 &       &       &       &  \\
327 & 276.743296 & $-33.606194$ & 13.874 & 13.244 & 13.065 &    20.670  & 4431 &  2.5  &$-0.31$&  0.3  &  \\
328 & 276.748212 & $-33.826157$ & 13.875 & 13.305 & 13.185 & $ -29.200$ & 4761 &       &       &       &  \\
329 & 276.494197 & $-33.811489$ & 13.876 & 13.340 & 13.159 &    10.821  & 4690 &       &       &       &  \\
330 & 276.905290 & $-33.690044$ & 13.879 & 13.243 & 13.114 & $ -21.543$ & 4542 &       &       &       &  \\
331 & 276.661029 & $-33.805126$ & 13.882 & 13.245 & 13.169 & $ -85.801$ & 4696 &       &       &       &  \\
332 & 276.725178 & $-33.922646$ & 13.894 & 13.301 & 13.118 &    74.659  & 4514 &       &       &       &  \\
333 & 276.803651 & $-33.785221$ & 13.895 & 13.274 & 13.161 & $-160.010$ & 4630 &       &       &       &  \\
334 & 276.573211 & $-33.613113$ & 13.897 & 13.384 & 13.189 &    30.030  & 4718 &       &       &       & *\\
335 & 276.754422 & $-33.812862$ & 13.899 & 13.284 & 13.177 & $  -3.150$ & 4666 &       &       &       &  \\
336 & 276.775634 & $-33.868172$ & 13.902 & 13.205 & 13.179 & $ -19.149$ & 4661 &       &       &       & *\\
337 & 276.609719 & $-33.899956$ & 13.910 & 13.284 & 13.186 &    86.522  & 4663 &       &       &       &  \\
338 & 276.802770 & $-33.607288$ & 13.910 & 13.270 & 13.182 & $ -31.855$ & 4651 &       &       &       & *\\
339 & 276.599163 & $-33.751358$ & 13.916 & 13.369 & 13.197 &    21.543  & 4682 &       &       &       &  \\
340 & 276.862496 & $-33.881008$ & 13.916 & 13.287 & 13.203 & $  -7.419$ & 4688 &       &       &       &  \\
342 & 276.619675 & $-33.905045$ & 13.925 & 13.343 & 13.216 &   123.234  & 4708 &       &       &       &  \\
343 & 276.694856 & $-33.562214$ & 13.926 & 13.257 & 13.149 &    69.723  & 4519 &       &       &       &  \\
344 & 276.586781 & $-33.832466$ & 13.931 & 13.301 & 13.158 &   214.266  & 4528 &       &       &       &  \\
347 & 276.736184 & $-33.875690$ & 13.949 & 13.316 & 13.213 & $ -45.041$ & 4624 &       &       &       &  \\
348 & 276.578552 & $-33.619892$ & 13.950 & 13.352 & 13.202 &     5.935  & 4600 &       &       &       &  \\
349 & 276.737919 & $-33.695145$ & 13.956 & 13.350 & 13.207 & $ -65.841$ & 4590 &       &       &       &  \\
350 & 276.456610 & $-33.796585$ & 13.957 & 13.347 & 13.225 & $-120.618$ & 4647 &       &       &       &  \\
351 & 276.564966 & $-33.635128$ & 13.960 & 13.423 & 13.248 &   128.623  & 4706 &  2.5  &$+0.17$&  1.2  &  \\
353 & 276.861593 & $-33.801018$ & 13.974 & 13.438 & 13.274 &    22.706  & 4728 &       &       &       &  \\
354 & 276.722013 & $-33.592499$ & 13.974 & 13.376 & 13.238 &    46.922  & 4630 &       &       &       &  \\
355 & 276.749304 & $-33.665966$ & 13.975 & 13.357 & 13.210 & $ -58.417$ & 4547 &       &       &       &  \\
356 & 276.762247 & $-33.786091$ & 13.982 & 13.460 & 13.272 &    20.393  & 4702 &       &       &       &  \\
357 & 276.785991 & $-33.896404$ & 13.985 & 13.341 & 13.224 & $ -84.393$ & 4553 &       &       &       &  \\
358 & 276.833052 & $-33.776608$ & 13.989 & 13.368 & 13.235 & $  -0.947$ & 4572 &       &       &       &  \\
359 & 276.433122 & $-33.740566$ & 13.989 & 13.411 & 13.292 & $ -32.267$ & 4753 &       &       &       &  \\
360 & 276.691675 & $-33.846764$ & 13.993 & 13.347 & 13.275 & $ -38.991$ & 4679 &       &       &       &  \\
361 & 276.705482 & $-33.740154$ & 13.996 & 13.411 & 13.280 &    37.993  & 4687 &       &       &       &  \\
362 & 276.617111 & $-33.760021$ & 13.996 & 13.410 & 13.278 &     7.110  & 4684 &       &       &       &  \\
363 & 276.493794 & $-33.799362$ & 13.996 & 13.417 & 13.299 &    30.066  & 4750 &       &       &       &  \\
364 & 276.697108 & $-33.596195$ & 13.996 & 13.460 & 13.297 &    49.761  & 4741 &       &       &       &  \\
365 & 276.575708 & $-33.577957$ & 14.000 & 13.428 & 13.271 &     0.930  & 4656 &       &       &       &  \\
366 & 276.663677 & $-33.647522$ & 14.002 & 13.400 & 13.306 &    12.679  & 4750 &       &       &       &  \\
367 & 276.779362 & $-33.567799$ & 14.002 & 13.382 & 13.217 &   188.004  & 4495 &       &       &       &  \\
368 & 276.871279 & $-33.792786$ & 14.006 & 13.428 & 13.318 & $ -35.508$ & 4764 &       &       &       &  \\
369 & 276.580553 & $-33.632629$ & 14.007 & 13.438 & 13.271 & $ -80.578$ & 4634 &       &       &       &  \\
370 & 276.711834 & $-33.703365$ & 14.013 & 13.394 & 13.306 &    48.722  & 4714 &       &       &       &  \\
371 & 276.743577 & $-33.898552$ & 14.014 & 13.444 & 13.263 & $ -93.685$ & 4581 &       &       &       & *\\
373 & 276.441719 & $-33.814545$ & 14.024 & 13.424 & 13.276 &     2.274  & 4601 &       &       &       &  \\
374 & 276.570897 & $-33.627617$ & 14.025 & 13.431 & 13.228 & $ -16.193$ & 4468 &       &       &       &  \\
376 & 276.747908 & $-33.773846$ & 14.027 & 13.360 & 13.322 & $ -31.917$ & 4718 &       &       &       &  \\
377 & 276.641959 & $-33.642696$ & 14.027 & 13.439 & 13.293 &     9.039  & 4638 &       &       &       & *\\
378 & 276.686341 & $-33.926971$ & 14.032 & 13.484 & 13.341 &   119.348  & 4758 &       &       &       &  \\
379 & 276.867112 & $-33.799599$ & 14.035 & 13.406 & 13.344 &    36.952  & 4755 &       &       &       &  \\
380 & 276.797940 & $-33.867813$ & 14.041 & 13.329 & 13.276 &    10.755  & 4542 &       &       &       &  \\
381 & 276.615896 & $-33.679054$ & 14.044 & 13.455 & 13.337 & $ -41.469$ & 4718 &       &       &       &  \\
382 & 276.642599 & $-33.707905$ & 14.051 & 13.414 & 13.319 &    29.732  & 4642 &       &       &       & *\\
383 & 276.552809 & $-33.793800$ & 14.052 & 13.441 & 13.273 &    58.636  & 4514 &       &       &       &  \\
384 & 276.767664 & $-33.790257$ & 14.054 & 13.476 & 13.288 & $ -40.420$ & 4542 &       &       &       &  \\
385 & 276.519644 & $-33.920509$ & 14.060 & 13.538 & 13.278 & $ -29.684$ & 4504 &       &       &       & *\\
386 & 276.778487 & $-33.904800$ & 14.063 & 13.482 & 13.386 &     7.124  & 4799 &       &       &       &  \\
387 & 276.738181 & $-33.939156$ & 14.068 & 13.462 & 13.390 &    93.169  & 4797 &       &       &       & *\\
388 & 276.529173 & $-33.642719$ & 14.071 & 13.505 & 13.331 & $-107.163$ & 4624 &       &       &       &  \\
389 & 276.805911 & $-33.751480$ & 14.078 & 13.399 & 13.380 &   179.763  & 4737 &       &       &       & *\\
390 & 276.654480 & $-33.594334$ & 14.081 & 13.422 & 13.341 & $  -8.853$ & 4621 &       &       &       &  \\
392 & 276.628139 & $-33.834789$ & 14.097 & 13.540 & 13.394 &     7.430  & 4726 &       &       &       & *\\
393 & 276.710695 & $-33.846008$ & 14.100 & 13.400 & 13.312 &    84.567  & 4484 &       &       &       &  \\
394 & 276.689541 & $-33.778378$ & 14.107 & 13.531 & 13.367 &    28.170  & 4616 &       &       &       &  \\
395 & 276.514699 & $-33.757675$ & 14.110 & 13.464 & 13.352 &   143.624  & 4572 &       &       &       &  \\
396 & 276.694978 & $-33.943527$ & 14.117 & 13.546 & 13.403 &    90.426  & 4689 &       &       &       &  \\
397 & 276.828208 & $-33.699005$ & 14.118 & 13.473 & 13.418 &    31.582  & 4731 &       &       &       &  \\
398 & 276.515313 & $-33.784389$ & 14.119 & 13.550 & 13.332 & $ -78.551$ & 4493 &       &       &       &  \\
399 & 276.634575 & $-33.901833$ & 14.126 & 13.579 & 13.433 & $ -52.671$ & 4755 &       &       &       &  \\
400 & 276.829987 & $-33.806454$ & 14.127 & 13.557 & 13.385 & $ -47.000$ & 4605 &       &       &       &  \\
401 & 276.744083 & $-33.730270$ & 14.129 & 13.516 & 13.358 & $ -64.671$ & 4530 &       &       &       &  \\
402 & 276.879971 & $-33.675831$ & 14.138 & 13.436 & 13.357 &    13.679  & 4500 &       &       &       &  \\
403 & 276.873784 & $-33.655762$ & 14.143 & 13.553 & 13.461 &   181.139  & 4785 &       &       &       &  \\
404 & 276.750878 & $-33.678814$ & 14.145 & 13.559 & 13.397 & $ -16.703$ & 4593 &       &       &       &  \\
406 & 276.773448 & $-33.816742$ & 14.159 & 13.602 & 13.453 &    89.280  & 4713 &       &       &       &  \\
407 & 276.538950 & $-33.857552$ & 14.167 & 13.626 & 13.426 & $ -59.382$ & 4617 &       &       &       &  \\
408 & 276.702749 & $-33.817944$ & 14.168 & 13.557 & 13.489 & $ -91.236$ & 4797 &       &       &       &  \\
409 & 276.578957 & $-33.705822$ & 14.171 & 13.622 & 13.446 &     0.475  & 4665 &       &       &       &  \\
410 & 276.753490 & $-33.738415$ & 14.173 & 13.552 & 13.448 &    24.100  & 4659 &       &       &       &  \\
411 & 276.836391 & $-33.686768$ & 14.174 & 13.593 & 13.453 &    21.788  & 4669 &       &       &       &  \\
412 & 276.501238 & $-33.637436$ & 14.174 & 13.595 & 13.412 & $-161.418$ & 4563 &       &       &       &  \\
413 & 276.662487 & $-33.626492$ & 14.179 & 13.651 & 13.492 &    50.701  & 4778 &       &       &       &  \\
414 & 276.628673 & $-33.566936$ & 14.185 & 13.516 & 13.421 & $  -5.715$ & 4555 &       &       &       &  \\
415 & 276.534514 & $-33.915627$ & 14.195 & 13.583 & 13.469 &    58.092  & 4660 &       &       &       &  \\
416 & 276.673953 & $-33.712120$ & 14.197 & 13.610 & 13.502 & $-111.533$ & 4751 &       &       &       &  \\
417 & 276.561145 & $-33.898586$ & 14.197 & 13.540 & 13.448 &    47.080  & 4592 &       &       &       &  \\
419 & 276.578991 & $-33.573586$ & 14.206 & 13.587 & 13.472 & $ -41.636$ & 4641 &       &       &       &  \\
420 & 276.687936 & $-33.576183$ & 14.212 & 13.551 & 13.441 & $ -69.762$ & 4535 &  2.5  &$+0.05$&  0.5  & *\\
422 & 276.852338 & $-33.874542$ & 14.215 & 13.578 & 13.489 & $ -20.986$ & 4650 &       &       &       &  \\
423 & 276.852706 & $-33.851723$ & 14.232 & 13.667 & 13.550 &    45.436  & 4782 &       &       &       &  \\
424 & 276.886130 & $-33.787621$ & 14.234 & 13.670 & 13.553 &   101.421  & 4785 &       &       &       &  \\
425 & 276.661265 & $-33.711018$ & 14.238 & 13.657 & 13.487 &    23.665  & 4587 &       &       &       &  \\
426 & 276.616342 & $-33.801567$ & 14.238 & 13.734 & 13.527 & $-107.099$ & 4704 &       &       &       &  \\
427 & 276.858769 & $-33.890724$ & 14.238 & 13.579 & 13.486 & $-152.559$ & 4575 &       &       &       &  \\
428 & 276.683750 & $-33.836472$ & 14.242 & 13.685 & 13.528 & $ -70.446$ & 4692 &       &       &       &  \\
429 & 276.794492 & $-33.592255$ & 14.245 & 13.644 & 13.565 &     3.931  & 4796 &       &       &       &  \\
430 & 276.767692 & $-33.907009$ & 14.247 & 13.673 & 13.511 & $-154.454$ & 4622 &       &       &       &  \\
431 & 276.561013 & $-33.621979$ & 14.248 & 13.682 & 13.531 &    82.186  & 4692 &       &       &       & *\\
433 & 276.801658 & $-33.837830$ & 14.256 & 13.632 & 13.536 & $ -60.959$ & 4670 &       &       &       & *\\
434 & 276.665024 & $-33.724117$ & 14.258 & 13.665 & 13.550 & $ -22.264$ & 4713 &       &       &       & *\\
435 & 276.768679 & $-33.857334$ & 14.259 & 13.664 & 13.559 &    99.095  & 4730 &       &       &       &  \\
436 & 276.493553 & $-33.646214$ & 14.263 & 13.713 & 13.588 & $ -50.196$ & 4822 &       &       &       & *\\
437 & 276.827726 & $-33.647717$ & 14.268 & 13.643 & 13.542 & $ -52.388$ & 4655 &       &       &       &  \\
438 & 276.594804 & $-33.845116$ & 14.271 & 13.742 & 13.515 &     5.719  & 4573 &       &       &       &  \\
440 & 276.631808 & $-33.564571$ & 14.278 & 13.711 & 13.603 &    14.329  & 4819 &       &       &       &  \\
441 & 276.700830 & $-33.637337$ & 14.281 & 13.799 & 13.534 & $ -18.383$ & 4598 &       &       &       &  \\
443 & 276.727795 & $-33.730831$ & 14.297 & 13.769 & 13.573 & $ -32.104$ & 4663 &       &       &       &  \\
445 & 276.472846 & $-33.808666$ & 14.310 & 13.670 & 13.574 &    73.543  & 4634 &  2.6  &$+0.25$&  0.7  &  \\
446 & 276.438307 & $-33.774590$ & 14.311 & 13.721 & 13.542 & $  -8.678$ & 4544 &       &       &       &  \\
447 & 276.777476 & $-33.775871$ & 14.316 & 13.739 & 13.634 & $ -81.331$ & 4786 &       &       &       &  \\
448 & 276.706230 & $-33.843826$ & 14.316 & 13.735 & 13.588 & $ -20.233$ & 4649 &       &       &       &  \\
449 & 276.798371 & $-33.735714$ & 14.320 & 13.855 & 13.631 &     5.772  & 4765 &       &       &       &  \\
451 & 276.887413 & $-33.852146$ & 14.324 & 13.738 & 13.636 & $-184.313$ & 4762 &  2.6  &$-1.05$&  0.8  &  \\
452 & 276.661034 & $-33.891323$ & 14.334 & 13.821 & 13.654 & $ -76.626$ & 4794 &  2.6  &$-0.99$&  0.7  &  \\
453 & 276.690616 & $-33.808006$ & 14.336 & 13.754 & 13.596 & $ -48.148$ & 4615 &       &       &       &  \\
454 & 276.698507 & $-33.939789$ & 14.343 & 13.800 & 13.657 &   115.364  & 4773 &       &       &       &  \\
455 & 276.845583 & $-33.661392$ & 14.344 & 13.746 & 13.667 &    65.842  & 4802 &       &       &       &  \\
456 & 276.593340 & $-33.672127$ & 14.349 & 13.759 & 13.642 & $ -23.056$ & 4719 &       &       &       &  \\
457 & 276.833423 & $-33.746063$ & 14.353 & 13.767 & 13.581 & $ -23.375$ & 4524 &       &       &       &  \\
459 & 276.615328 & $-33.576347$ & 14.368 & 13.740 & 13.606 &    49.302  & 4561 &       &       &       &  \\
460 & 276.722855 & $-33.943268$ & 14.391 & 13.838 & 13.708 & $ -73.249$ & 4781 &       &       &       &  \\
461 & 276.834942 & $-33.635342$ & 14.396 & 13.863 & 13.666 & $ -19.355$ & 4643 &  2.7  &$-1.50$&  0.8  &  \\
462 & 276.789630 & $-33.841022$ & 14.400 & 13.855 & 13.727 &     4.985  & 4813 &       &       &       &  \\
463 & 276.467867 & $-33.749195$ & 14.400 & 13.783 & 13.650 & $-172.922$ & 4595 &       &       &       &  \\
464 & 276.657623 & $-33.651539$ & 14.406 & 13.786 & 13.632 & $ -66.065$ & 4526 &       &       &       &  \\
465 & 276.631871 & $-33.704411$ & 14.417 & 13.879 & 13.739 & $-121.778$ & 4806 &       &       &       &  \\
466 & 276.861515 & $-33.729511$ & 14.417 & 13.903 & 13.742 & $ -59.983$ & 4806 &       &       &       &  \\
467 & 276.742712 & $-33.819698$ & 14.422 & 13.889 & 13.745 & $  -9.504$ & 4802 &       &       &       &  \\
468 & 276.511268 & $-33.792122$ & 14.422 & 13.791 & 13.707 &    56.092  & 4696 &       &       &       &  \\
469 & 276.581345 & $-33.660347$ & 14.425 & 13.834 & 13.711 & $ -44.438$ & 4699 &       &       &       &  \\
471 & 276.439088 & $-33.761520$ & 14.447 & 13.839 & 13.738 &    35.018  & 4717 &       &       &       &  \\
472 & 276.823834 & $-33.629124$ & 14.464 & 13.925 & 13.731 &   131.626  & 4635 &       &       &       &  \\
473 & 276.670917 & $-33.612740$ & 14.466 & 13.942 & 13.796 &    24.364  & 4833 &       &       &       & *\\
474 & 276.547004 & $-33.669544$ & 14.475 & 13.910 & 13.772 & $ -17.630$ & 4733 &       &       &       &  \\
476 & 276.661846 & $-33.655987$ & 14.491 & 13.900 & 13.829 & $ -29.627$ & 4859 &       &       &       &  \\
477 & 276.853016 & $-33.748569$ & 14.496 & 13.867 & 13.809 & $ -75.148$ & 4768 &  2.7  &$-1.09$&  0.7  &  \\
478 & 276.533779 & $-33.846905$ & 14.501 & 13.982 & 13.812 & $ -38.002$ & 4772 &       &       &       &  \\
481 & 276.777363 & $-33.671219$ & 14.502 & 13.896 & 13.736 & $ -11.567$ & 4543 &       &       &       & *\\
482 & 276.573957 & $-33.858578$ & 14.505 & 14.030 & 13.800 & $-154.032$ & 4722 &  2.7  &$-0.78$&  0.9  &  \\
483 & 276.603613 & $-33.823669$ & 14.506 & 13.932 & 13.766 &    31.359  & 4618 &       &       &       &  \\
484 & 276.594676 & $-33.571365$ & 14.507 & 13.849 & 13.829 & $  -6.386$ & 4810 &       &       &       & *\\
485 & 276.530138 & $-33.766701$ & 14.511 & 13.876 & 13.792 &   129.335  & 4684 &       &       &       &  \\
486 & 276.865712 & $-33.679306$ & 14.515 & 13.996 & 13.838 & $ -56.181$ & 4801 &       &       &       &  \\
487 & 276.849420 & $-33.771736$ & 14.519 & 13.847 & 13.786 & $-141.982$ & 4631 &       &       &       &  \\
488 & 276.598639 & $-33.744186$ & 14.521 & 13.967 & 13.832 & $  -3.129$ & 4771 &       &       &       &  \\
489 & 276.875366 & $-33.742172$ & 14.521 & 13.810 & 13.769 &    73.034  & 4577 &       &       &       &  \\
490 & 276.712242 & $-33.794239$ & 14.531 & 13.928 & 13.856 & $  -2.514$ & 4810 &       &       &       &  \\
491 & 276.751951 & $-33.710484$ & 14.534 & 13.883 & 13.786 & $ -47.811$ & 4593 &       &       &       &  \\
493 & 276.631945 & $-33.662464$ & 14.538 & 13.956 & 13.862 &    54.638  & 4813 &  2.7  &$-1.12$&  0.9  & *\\
494 & 276.572343 & $-33.570671$ & 14.542 & 13.939 & 13.812 & $ -28.906$ & 4653 &       &       &       &  \\
495 & 276.548845 & $-33.833427$ & 14.549 & 13.924 & 13.793 &    34.963  & 4575 &       &       &       &  \\
496 & 276.765063 & $-33.868202$ & 14.555 & 13.990 & 13.859 &    77.859  & 4742 &       &       &       &  \\
497 & 276.830924 & $-33.896267$ & 14.555 & 13.988 & 13.816 & $ -30.904$ & 4612 &       &       &       &  \\
498 & 276.771491 & $-33.669971$ & 14.559 & 13.951 & 13.817 & $ -23.933$ & 4610 &       &       &       &  \\
499 & 276.633311 & $-33.762764$ & 14.567 & 13.899 & 13.838 &    31.087  & 4650 &       &       &       &  \\
500 & 276.602649 & $-33.731819$ & 14.567 & 14.041 & 13.883 & $  -1.571$ & 4787 &       &       &       &  \\
501 & 276.858813 & $-33.772926$ & 14.568 & 13.883 & 13.878 &   174.776  & 4759 &       &       &       &  \\
503 & 276.436893 & $-33.808731$ & 14.580 & 13.968 & 13.852 & $ -36.423$ & 4659 &       &       &       &  \\
504 & 276.481768 & $-33.625542$ & 14.581 & 13.971 & 13.872 & $ -67.245$ & 4718 &       &       &       &  \\
505 & 276.537031 & $-33.778503$ & 14.584 & 14.036 & 13.893 & $  -9.908$ & 4767 &       &       &       &  \\
506 & 276.502711 & $-33.686172$ & 14.585 & 13.967 & 13.838 & $ -10.462$ & 4604 &       &       &       &  \\
507 & 276.736185 & $-33.691521$ & 14.591 & 13.961 & 13.833 & $ -87.751$ & 4566 &       &       &       &  \\
508 & 276.761504 & $-33.739830$ & 14.594 & 13.971 & 13.862 & $-231.533$ & 4638 &  2.7  &$-0.71$&  0.9  &  \\
509 & 276.856836 & $-33.670521$ & 14.596 & 13.998 & 13.898 & $-150.306$ & 4737 &  2.7  &$-0.76$&  0.7  &  \\
511 & 276.586662 & $-33.710934$ & 14.606 & 13.974 & 13.869 &   106.628  & 4629 &       &       &       &  \\
512 & 276.620755 & $-33.646839$ & 14.620 & 14.124 & 13.932 & $ -19.865$ & 4776 &       &       &       &  \\
513 & 276.868502 & $-33.816536$ & 14.620 & 14.136 & 13.959 &    39.173  & 4851 &       &       &       &  \\
514 & 276.727030 & $-33.799290$ & 14.624 & 14.014 & 13.922 &    42.665  & 4727 &       &       &       & *\\
\end{longtable}
\tablefoot{Meaning of the columns: (1): Target number in this programme; (2):
Right ascension from 2MASS; (3): Declination from 2MASS; (4), (5), and (6):
2MASS $J$-, $H$-, and $K$-magnitude; (7): Heliocentric radial velocity; (8):
Effective temperature; (9): Surface gravity $\log g$; (10): Metallicity; (11):
Li abundance; (12): Flag for foreground stars.}
\label{tabappend}
}

\section{Discussion}

\subsection{The frequency of Li rich stars}

\begin{figure}
  \centering
  \includegraphics[width=\linewidth,bb=83 366 540 699,clip]{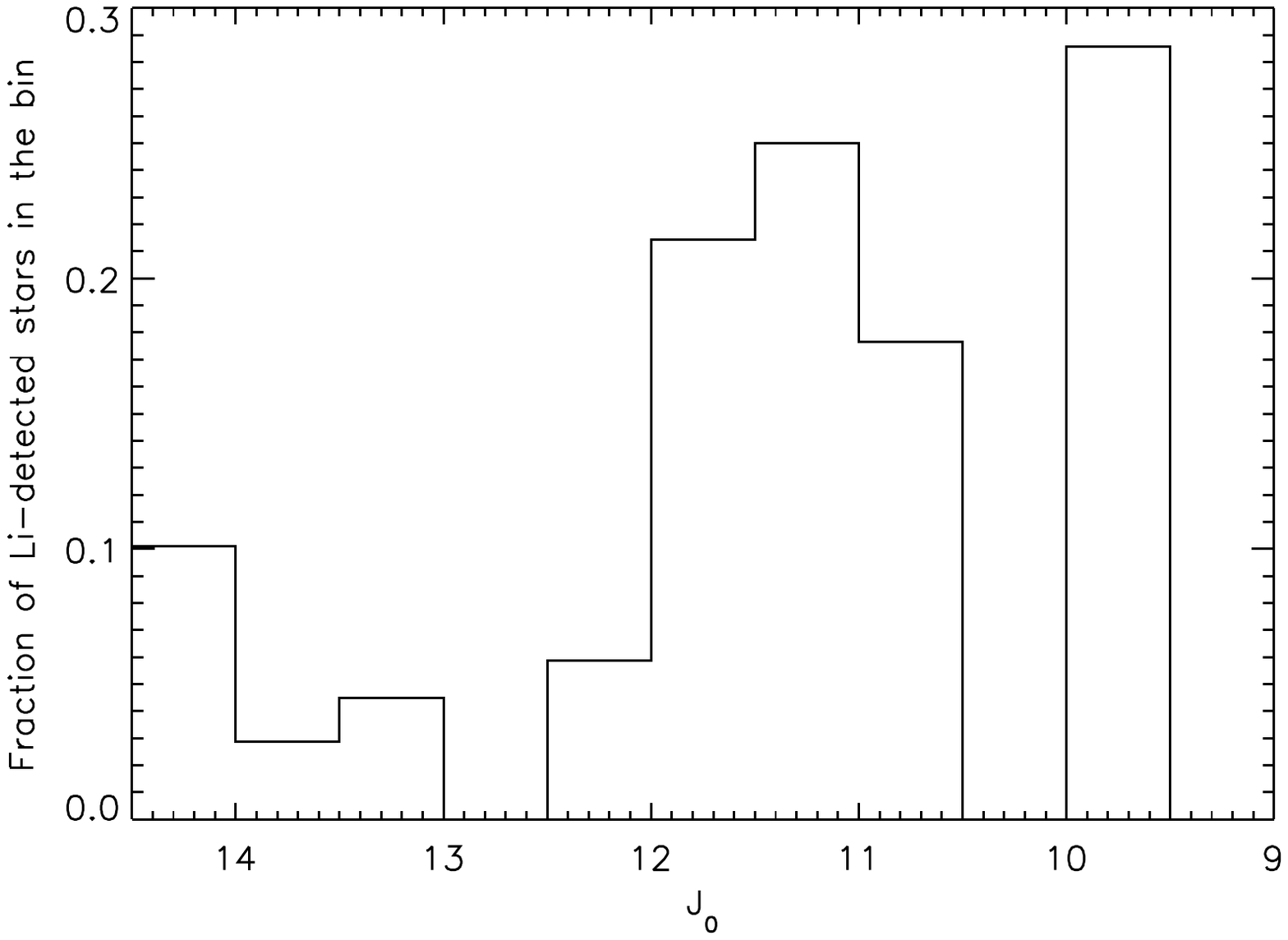}
  \caption{Fraction of Li-detected stars as a function of $J_0$.}
  \label{Lifract}
\end{figure}

The three most Li-rich stars in our sample are located on the upper RGB,
clearly above the bump and the two RCs. The faintest of our Li-rich stars
(\#123), at a $K$-magnitude of 11\fm61, is more than one magnitude brighter
than the bright RC (peak at $K\sim12\fm65$). Taking into account the
$K$-magnitude spread of the RCs of 0\fm22 (David Nataf, private communication),
this means that the faintest Li-rich star is brighter than the bright RC peak
by 4.7 standard deviations. Hence, even considering the error in distance caused
by the spread within the Bulge, it is extremely unlikely that the Li-rich stars
are related to either the RGB bump or clump. On the other hand, the
Li-{\em detected} stars are scattered all along the RGB, from below the bump to
the tip. Neither at the bump nor at the RCs do we see an overdensity of
Li-detected stars. In contrast, the {\em fraction} of Li-detected stars on the
upper RGB ($J_0 < 12\fm0$) is 18.8\%, much higher than below this limit
(5.4\%). This important result of the present study is illustrated in
Fig.~\ref{Lifract}, which shows a histogram of the fraction of Li-detected
stars as a function of $J_0$. The fraction of Li-detected stars also increases
as a function of $(J-K_s)_0$, with a local maximum of 26.3\% between
$(J-K_s)_0=0.8$ and 0.9.

Recently, U07 identified four AGB stars in the PG3 field with Li abundance
of $\log\epsilon({\rm Li})=0.8$, 0.8, 1.1, and 2.0, among a sample of 27
long-period AGB variables. The fraction of {\em upper} RGB stars with
detectable Li line is similar to that found among AGB stars, accordingly it is
possible that the Li-rich AGB stars inherited their Li from the preceding RGB
phase. Indeed, the three Li-rich stars identified in the present study could be
early AGB stars instead of RGB stars because they are brighter than the RC.
Because it is impossible to separate RGB and early AGB stars by means of our
photometric and spectroscopic data alone, asteroseismological methods would be
needed to better define their precise evolutionary state.

\subsection{Mass loss from Li-rich stars?}

\begin{figure}
  \centering
  \includegraphics[width=\linewidth,bb=71 373 538 699,clip]{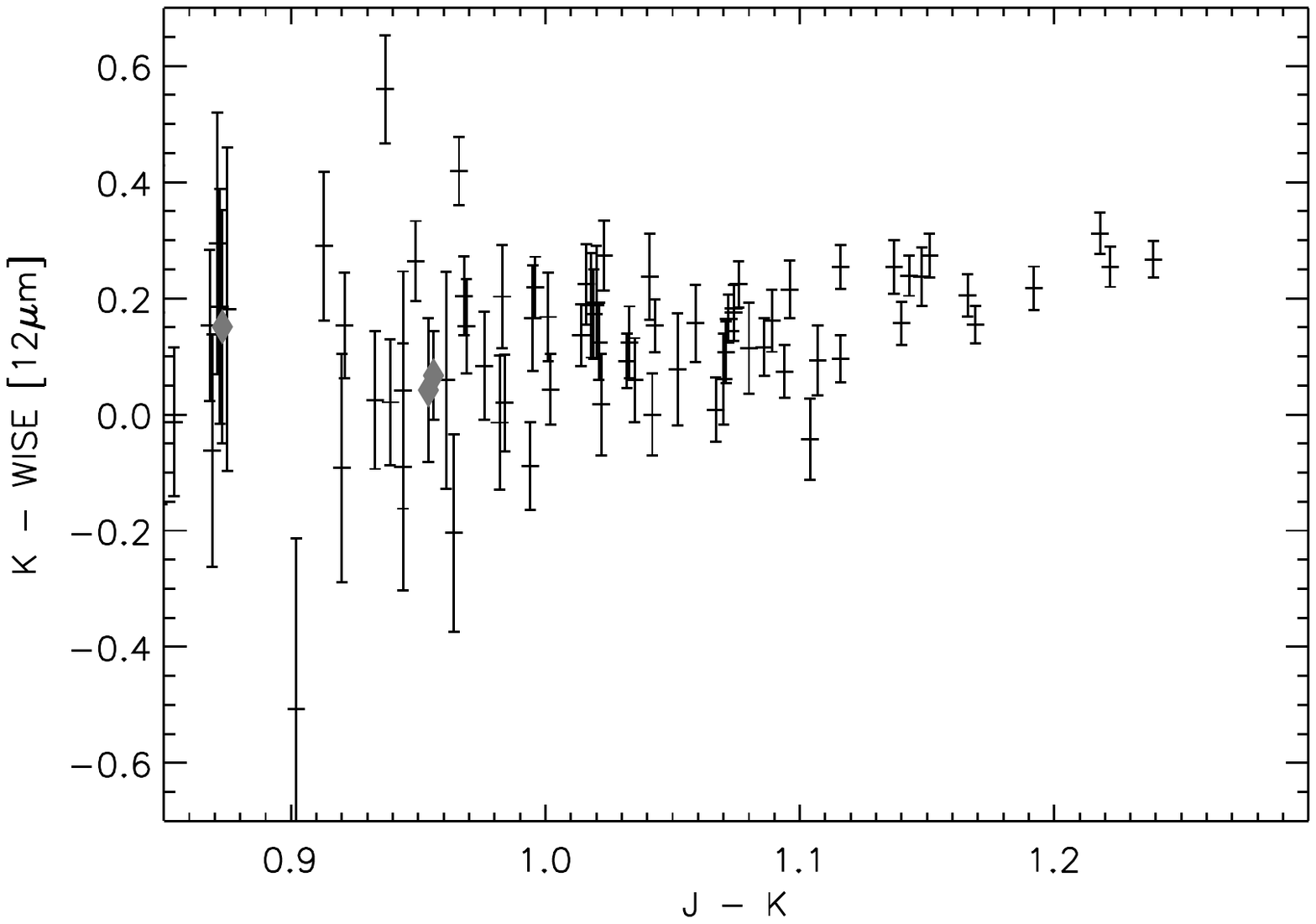}
  \caption{$K - \rm{WISE} [12{\rm \mu m}]$ vs.\ $J-K$ colour-colour diagram of
    our sample stars. The grey diamond symbols represent the three Li-rich
    stars.}
  \label{WISE_CCD}
\end{figure}

It has been speculated that the Li enrichment in K-type giants is accompanied
by a mass-loss episode \citep{dlR96,dlR97}, which has, however, later been
questioned \citep{FW98,Jas99}. To investigate if there is enhanced
{\em dust} mass-loss from our target stars, we cross-identified them with
sources detected by the WISE space observatory \citep{Wri10}. Within a search
radius of 1\farcs2 we found counterparts for 333 of our targets in the WISE
catalogue\footnote{http://irsa.ipac.caltech.edu/}. We inspected $J-K$ vs.\
$K - \rm{WISE}~[3.4{\rm \mu m}]$, $[4.6{\rm \mu m}]$, and $[12{\rm \mu m}]$
colour-colour diagrams of these stars. The 22\,${\rm \mu m}$ WISE band proved to
have too low a sensitivity to reliable detect more than just the very brightest
of our targets. A version of the $J-K$ vs.\ $K - \rm{WISE}~[12{\rm \mu m}]$
colour-colour diagram is displayed in Fig.~\ref{WISE_CCD}. The x-axis range was
restricted to the reddest part of the sample, because at the bluer (and
hence fainter) end of the sample the noise level is very high. The few red
outliers in the colour-colour diagrams were identified in the 2MASS images to
be visual pairs or triples that are resolved by 2MASS, but are probably
unresolved in the WISE observations. The three Li-rich stars have very
inconspicuous $K - \rm{WISE}$ colours, strongly suggesting that they do not
suffer enhanced dusty mass-loss. For comparison, the Miras investigated by U07,
which suffer a total mass-loss rate of $10^{-8} M_{\odot} {\rm yr}^{-1}$ or more,
have $K - [12{\rm \mu m}]$ colours of 1.17 or redder.

To search also for enhanced {\em gas} mass-loss, we inspected the spectra of
the Li-rich stars around the H$\alpha$ line in comparison with otherwise
similar Li-poor stars, namely \#051 and \#131. Gas mass-loss would be
detectable through blue-shifted asymmetries in the H$\alpha$ line profile, see
e.g.\ \citet{Mes09} for a sensitive search for gas mass-loss in red giants. The
most Li-rich sample star (\#042) has a slightly broader H$\alpha$ profile than
the comparison stars, but no asymmetry. The other two Li-rich stars have a
symmetric profile as well. We conclude from this that within the limits of
resolution and S/N of our spectra, the Li-rich stars do not suffer enhanced
loss of either dust nor gas, confirming the conclusions of \citet{FW98} and
\citet{Jas99}.

\subsection{A trend of Li abundance with temperature}

\begin{figure}
  \centering
  \includegraphics[width=\linewidth,bb=84 367 536 698,clip]{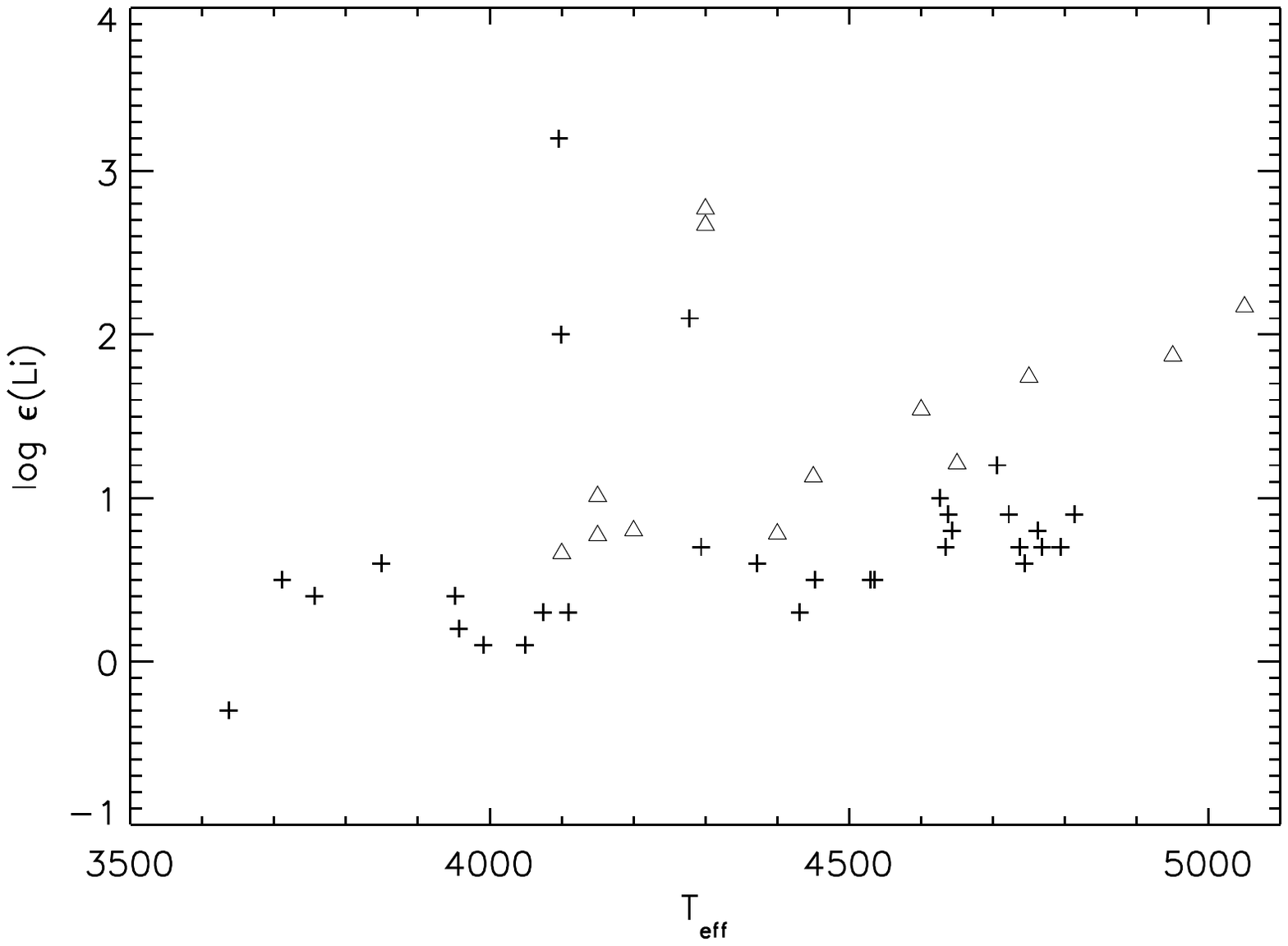}
  \caption{Lithium abundance as a function of effective temperatures of RGB
    stars. The data points from our study are plotted as plus symbols, while
    the open triangles represent the LTE measurements from \citet{Gon09}.}
  \label{ALiTeff}
\end{figure}

Recently, a study of Li-rich giants in the GB was presented by \citet{Gon09}.
Among a sample of 417 stars, chosen from a narrow brightness range located
about $0\fm7$ above the expected RC, they found 13 stars with detectable Li
line. \citet{Gon09} concluded that it is unlikely that the Li-rich stars are
connected to either the RGB bump or the RC if they are genuine GB stars at a
mean distance of 8\,kpc. Although the authors' sample covers only a narrow
luminosity range along the RGB, their results agree with ours. \citet{Gon09}
report a decrease of the Li abundance in the Li-detected stars with decreasing
temperature in their sample, which again agrees with our result. A comparison
with their LTE results is presented in Fig.~\ref{ALiTeff}. Our sample extends
the trend found by \citet{Gon09} to lower temperatures. Just as in
\citet{Gon09}, our Li-rich stars also deviate from this trend, and fall in
a similar temperature range at $T_{\rm eff} = \sim 4100 - 4300$\,K. One
difference to our study is, however, that \citet{Gon09} find significantly
higher Li-abundances at the hot end of their sample ($T_{\rm eff} \geq 4600$\,K)
than we do. This difference is mostly caused by four stars in the \citet{Gon09}
sample at $T_{\rm eff} \geq 4600$\,K and $\log\epsilon({\rm Li})\geq1.5$. Those
stars also have a lower surface gravity ($\log g = 1.9 - 2.1$) than we estimate
for our sample stars in this temperature range ($\log g = 2.3 - 2.7$). Thus,
there could be a difference in evolutionary state between these groups of
stars, also in terms of Li abundance, but we are not sure if this fully
accounts for the found difference.

Furthermore, \citet{Gon09} also carefully examined whether the Li-rich stars are
special in any other way, but did not find a difference to the Li-normal stars.
In a similar attempt, we inspected the spectra of the Li-rich and Li-detected
stars in our sample in comparison with Li-poor stars to search for
peculiarities, but we did not find any. We only found two sample stars with
increased strength in the BaII 649.9\,nm and the LaII 652.9\,nm lines. They
probably belong to the so-called barium star class, collecting members of
binary systems that owe their enhanced s-element abundances to a mass-transfer
episode from a more massive AGB companion. These will be discussed in more
detail in a forthcoming paper. The Li-rich and Li-detected stars are perfectly
normal in their BaII and LaII line strengths, which certifies that their Li
overabundance is not the result of mass transfer of Li-enriched matter from a
binary (former AGB) companion. Finally, \citet{Gon09} also conclude that their
Li-rich stars do not suffer enhanced mass-loss, based 2MASS $JHK$ photometry.

The three stars with the highest Li abundance in our sample fall nicely within
the sequence of Li-rich thick disk giants identified by
\citet[][their Fig.~4]{Mon11}. Recently, \citet{Kumar11} also identified
Li-rich giants in the Galactic disk and ascribed them to the red clump, i.e.\
these are stars after the helium core flash. We found no accumulation of
Li-rich stars close to the RCs in our sample. This could be related to a
difference in mass between the stars in our study and that of \citet{Kumar11}:
While there are mostly low-mass stars with a narrow distribution around
$1 M_{\sun}$ in our sample, the disk sample of \citet{Kumar11} likely also
contains stars of higher mass, whose Li abundance may evolve differently. The
thick disk sample of \citet{Mon11} likely contains mostly stars of low mass as
well, although the authors identify one star that could have a mass of up to
4\,$M_{\sun}$, albeit with large uncertainty. The agreement between our results
and those of \citet{Mon11} is probably related to a similar mass spectrum of
the samples.

\section{Conclusions}

There are now several studies that show that at low to intermediate masses
Li-rich stars can be found all along the RGB, not necessarily connected to a
particular phase of the giant branch evolution \citep{Gon09,Mon11,Alc11,Ruch11}.
This means that either the Li enrichment process is stochastic in nature and can
happen at any time on the RGB, or that Li enrichment starts at some point on
the lower RGB and may take a long time to make Li-rich K giants. The activation
of mixing events of variable duration and transport rates below the convective
envelope is conceivable for the Li enrichment \citep{Palm11}. The lack of
a well-defined Li-rich phase on the RGB, as now shown by several recent
studies, contradicts predictions by CB00, who propose that a Li-rich
phase should occur at the bump of the RGB in low-mass stars, such as those that
we have in our sample. The existence of such a Li-rich phase has to be
questioned.

\begin{acknowledgements}
  TL acknowledges support from the Austrian Science Fund (FWF) under projects
  P~21988-N16 and P~23737-N16, and SU under project P~22911-N16. BA thanks for
  the support from contract ASI-INAF I/009/10/0. This publication makes use of
  data products from the Two Micron All Sky Survey, which is a joint project of
  the University of Massachusetts and the Infrared Processing and Analysis
  Center/California Institute of Technology, funded by the National Aeronautics
  and Space Administration and the National Science Foundation. This
  publication makes use of data products from the Wide-field Infrared Survey
  Explorer, which is a joint project of the University of California, Los
  Angeles, and the Jet Propulsion Laboratory/California Institute of
  Technology, funded by the National Aeronautics and Space Administration.
\end{acknowledgements}


%

\Online

\begin{appendix}

\section{The $(J-K_S) - T_{\rm{eff}}$ calibration from COMARCS models}
\label{JKTeffsect}

\begin{figure}
  \centering
  \includegraphics[width=\linewidth,bb=71 366 545 699,clip]{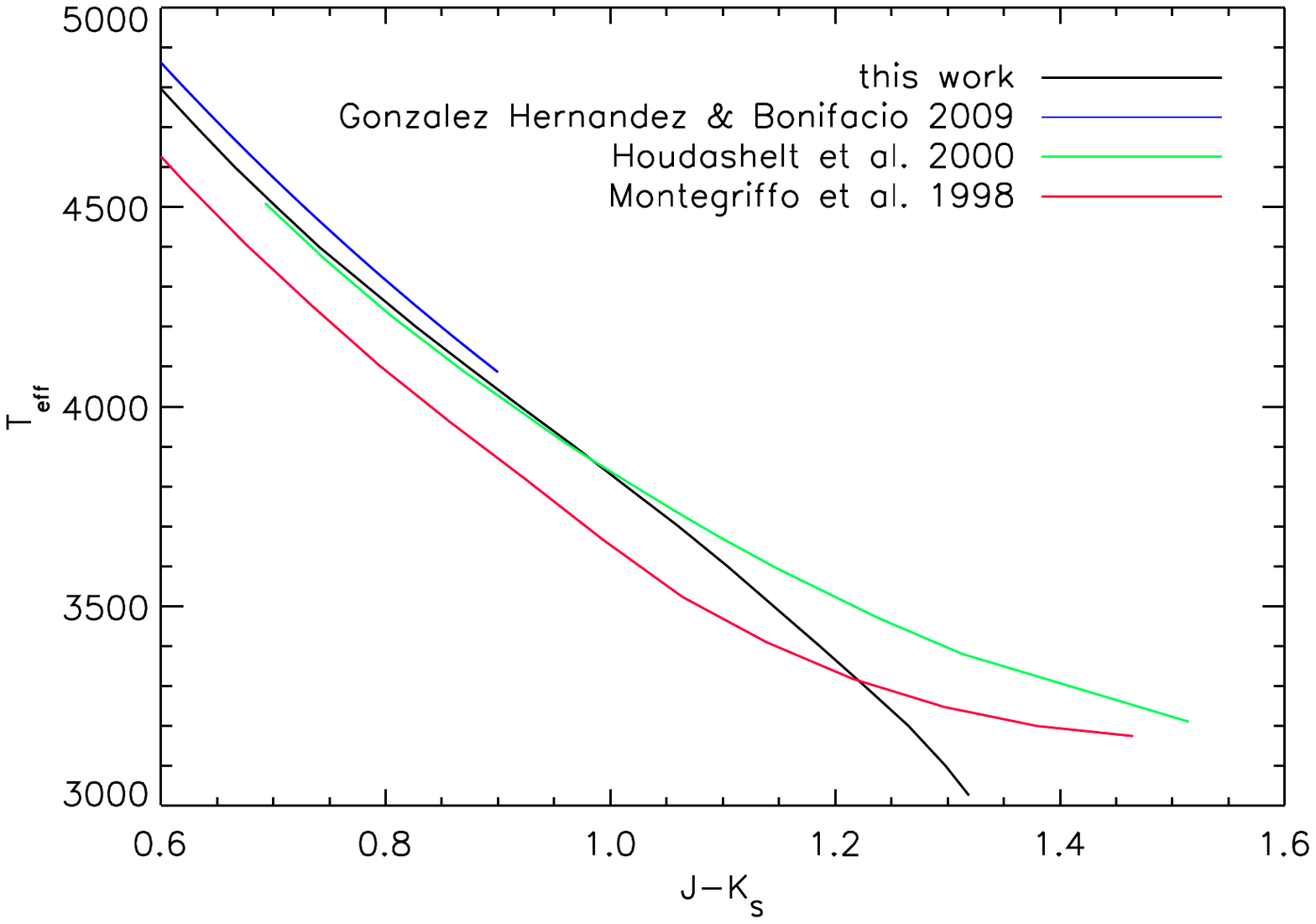}
  \caption{Comparison between different $(J-K_S) - T_{\rm{eff}}$ calibrations as
    identified in the legend.}
  \label{JKTefffig}
   \end{figure}

In Table~\ref{JKTefftab} and Fig.~\ref{JKTefffig} we present an effective
temperature calibration for the 2MASS $J-K_s$ colour based on hydrostatic
COMARCS atmospheres and the COMA spectral synthesis package
\citep[see][]{Ari09}. The same input opacities were used for the construction
of the models and the subsequent radiative transfer calculations. A list
including most of the incorporated molecular species and data can be found in
\citet{MA09}. The calculations are consistent with the spectral synthesis
applied in this work to obtain metallicities and Li abundances.

For the presented models we assumed a microturbulent velocity of
$\xi=2.5$\,km\,s$^{-1}$, solar abundances from \citet{GS98}, with revisions
from \citet{Caf08,Caf09} and surface gravities based on a typical solar mass
and abundance evolutionary sequence \citep{Mar08}. Compared to the effective
temperature, the influence of metallicity and surface gravity on the
predicted $J-K_s$ colours is fairly small. We expect a maximum uncertainty of
$\pm50$\,K for our programme stars owing to the variation of these quantities.

A comparison between different $(J-K_S) - T_{\rm{eff}}$ calibrations is shown in
Fig.~\ref{JKTefffig}. Our calibration (black graph) is compared to those of
\citet[][blue graph]{GHB09}, \citet[][green graph]{Hou00}, and of
\citet[][red graph]{Monte98}. The relations of \citet{Hou00} and \citet{Monte98}
were transformed from the Bessel \& Brett \citep{BB88} and ESO \citep{vdB96}
systems, respectively, to the 2MASS system using the relations of
\citet{Carpenter}. Solar metallicity ([Fe/H]=0.0) was assumed for the relation
of \citet{GHB09}, but the dependence on me\-tal\-li\-ci\-ty is weak. This last
relation is only valid up to $J-K_S=0.9$, but an extrapolation to redder values
agrees well with the other relations. The relation of \citet{Monte98} results
in lower temperatures than the others by about 200\,K in a broad range of
$J-K_S$ values, whereas the difference between the other three relations, in
the region of overlap, is small. Our relation agrees particularly well with
that of \citet{Hou00} up to $J-K_S \sim 1.1$, but it subsequently diverges.

\begin{table}
\caption{$(J-K_S) - T_{\rm{eff}}$ calibration from COMARCS models}
\label{JKTefftab}      
\centering                          
\begin{tabular}{ccrccc}        
\hline\hline                 
$T_{\rm{eff}}$ & $J-K_S$ & $\log g$ & BC(J) & BC(H) & BC(K) \\    
\hline                        
5800 & 0.343 &  4.24 & 1.108 & 1.404 & 1.451 \\
5700 & 0.363 &  4.12 & 1.141 & 1.453 & 1.504 \\
5600 & 0.384 &  4.05 & 1.174 & 1.502 & 1.558 \\
5500 & 0.407 &  4.01 & 1.205 & 1.552 & 1.612 \\
5400 & 0.430 &  3.98 & 1.237 & 1.603 & 1.667 \\
5300 & 0.454 &  3.95 & 1.268 & 1.653 & 1.722 \\
5200 & 0.481 &  3.92 & 1.298 & 1.705 & 1.779 \\
5100 & 0.509 &  3.88 & 1.327 & 1.757 & 1.836 \\
5000 & 0.538 &  3.82 & 1.356 & 1.810 & 1.894 \\
4900 & 0.569 &  3.70 & 1.385 & 1.864 & 1.954 \\
4800 & 0.599 &  3.35 & 1.417 & 1.919 & 2.016 \\
4700 & 0.632 &  3.03 & 1.447 & 1.973 & 2.079 \\
4600 & 0.666 &  2.77 & 1.477 & 2.028 & 2.143 \\
4500 & 0.703 &  2.52 & 1.505 & 2.083 & 2.208 \\
4400 & 0.741 &  2.31 & 1.532 & 2.138 & 2.273 \\
4300 & 0.784 &  2.11 & 1.557 & 2.193 & 2.341 \\
4200 & 0.827 &  1.91 & 1.582 & 2.247 & 2.409 \\
4100 & 0.873 &  1.72 & 1.606 & 2.302 & 2.479 \\
4000 & 0.920 &  1.54 & 1.630 & 2.357 & 2.550 \\
3900 & 0.968 &  1.35 & 1.656 & 2.412 & 2.624 \\
3800 & 1.014 &  1.17 & 1.685 & 2.469 & 2.699 \\
3701 & 1.060 &  1.00 & 1.716 & 2.526 & 2.776 \\
3600 & 1.104 &  0.82 & 1.752 & 2.586 & 2.856 \\
3500 & 1.145 &  0.67 & 1.792 & 2.646 & 2.937 \\
3400 & 1.186 &  0.49 & 1.835 & 2.705 & 3.021 \\
3300 & 1.226 &  0.33 & 1.879 & 2.761 & 3.105 \\
3200 & 1.265 &  0.17 & 1.923 & 2.817 & 3.188 \\
3100 & 1.298 &  0.03 & 1.966 & 2.873 & 3.264 \\
3025 & 1.319 & --0.11& 1.998 & 2.912 & 3.317 \\
\hline                                   
\end{tabular}
\end{table}

\end{appendix}

\end{document}